\documentclass[
 prr,reprint,
 amsmath,amssymb,
 aps,
floatfix,
dvipsnames,
longbibliography]{revtex4-2}

\usepackage[normalem]{ulem}
\usepackage{subfiles}
\usepackage{natbib}
\usepackage{graphicx}
\usepackage{dcolumn}
\usepackage{bm}
\usepackage{hyperref}

\usepackage{xcolor}
\newcommand{\michele}[1]{\textcolor{Black}{#1}}
\newcommand{\miha}[1]{\textcolor{Black}{#1}}

\newcommand{\hig}[1]{\textcolor{Black}{#1}}

\begin{document}

\preprint{APS/123-QED}

\title{\miha{Quantum R\'enyi entropy by optimal thermodynamic integration paths}}

\author{Miha Srdinšek\textsuperscript{1,2,3}}
\email{miha.srdinsek@upmc.fr}
\author{Michele Casula\textsuperscript{2}}%

\author{Rodolphe Vuilleumier\textsuperscript{3}}%
\affiliation{
 \textsuperscript{1}Institut des sciences du calcul et des données (ISCD), Sorbonne Universit\'e, 4 Place Jussieu, 75005 Paris.\\
 \textsuperscript{2}Sorbonne Universit\'e, Institut de minéralogie, de physique des matériaux et de cosmochimie (IMPMC),  CNRS UMR 7590, MNHM, 4 Place Jussieu, 75005 Paris.\\
 \textsuperscript{3}Processus d’Activation Sélectif par Transfert d’Energie Uni-électronique ou Radiative (PASTEUR), CNRS UMR 8640, D\'epartement de Chimie, \'Ecole Normale Superieure, 24 rue Lhomond, 75005 Paris.
 }

\date{\today}

\begin{abstract}
Despite being a well-established operational approach to quantify entanglement, R\'enyi entropy calculations have been plagued by their computational complexity. We introduce here a theoretical framework based on an optimal thermodynamic integration scheme, where the R\'enyi entropy can be efficiently evaluated using regularizing paths. This approach avoids slowly convergent fluctuating contributions and leads to low-variance estimates. In this way, large system sizes and high levels of entanglement in model or first-principles Hamiltonians are within our reach. We demonstrate it in the one-dimensional quantum Ising model and perform the evaluation of entanglement entropy in the formic acid dimer, by discovering that its two shared protons are entangled even above room temperature.
\end{abstract}

\maketitle

\emph{Introduction.}{---}
\michele{Measuring the entanglement of a quantum state or the entropy of a quantum system at thermal equilibrium has always been a challenge. With the aim of achieving this goal, several methods have been proposed so far\cite{Horodecki2009,Hastings2010,Humeniuk2012,Alba2017,Demidio2020,White2004,Do2013,Tony2010,Luitz2014}. One of the most promising approaches is based upon the evaluation of the quantum R\'enyi entropy. For the subsystem $A$ of a quantum system, it}\miha{is defined as\cite{Horodecki2009, Preskill2016, Cover1938, Horodecki1996}}
\begin{eqnarray}
S_A^{\alpha} = \frac{1}{1-\alpha}\log\frac{\text{Tr}\rho_A^{\alpha}}{ [\text{Tr}\rho]^{\alpha}},
\label{eq:RenyiEntropyDef}
\end{eqnarray}
where $\rho$ and $\rho_A$ are density matrices of the full system and of its subsystem $A$, respectively, with $\alpha\in\mathbb{R}_{>0}\setminus\{1\}$. \miha{When $\rho_A$ equals the full density matrix $\rho$, $S_A^{\alpha}$ serves as a very general signature of thermodynamic phase transitions. Moreover, when a smaller subsystem is considered, it can detect quantum phase transitions\cite{Vidal2003, Calabrese2004, Cirac2008, Flammia2009, Metlitski2009, Hastings2010, Singh2011, Romera2011, Humeniuk2012, Herdman2014} occurring at zero temperature, and quantifies the entanglement of the ground state. Together with its derivatives, like mutual information\cite{Cirac2008}, topological entropy\cite{Kitaev2006} or simply entropic inequalities\cite{Horodecki1996, Horodecki2009, Islam2015}, it can be used to classify different ground states, topological phases, extract critical exponents and study thermalization under unitary time evolution\cite{DAlessio2016, Nandkishore2015, Alet2018}.}

Despite being a fundamental proxy to understand the thermodynamics of quantum systems, the \miha{R\'enyi} entropy is not measurable in experimental setups, apart from rare successes\cite{Islam2015, Brydges2020}. The same holds for the R\'enyi entropy evaluation via analytical treatments or numerical methods, such as density matrix renormalization group (DMRG) and stochastic sampling frameworks\cite{Hastings2010, Humeniuk2012, Luitz2014, Herdman2014a, Singh2011, Alba2017,  Demidio2020, Zhao2022}. In the former case, it is limited mostly to integrable models\cite{Calabrese2004, Fan2004, Refael2004, Franchini2007, Bertini2019}, while in the latter, it is limited to low-dimensional systems with sufficiently low entanglement\cite{Schuch2008, Vidal2008, Carleo2017}. Quantifying the R\'enyi entropy in more general complex systems has remained so far an unattainable task, hampered by the exponentially high energy barriers in stochastic sampling frameworks.

Of particular interest are hydrogen-rich materials, like liquid water\cite{Ceriotti2013}, where quantum effects arise due to the light mass of hydrogen. It has been shown that nuclear quantum effects pilot phase transitions in these systems, leading, for instance, to phase VIII of water ice and to the superconducting phases of hydrides, such as LaH$_{10}$\cite{Drozdov2019}, YH$_n$\cite{Kong2021}, and H$_3$S\cite{Drozdov2015}. For the same reasons, entanglement is supposed to play a role also in biochemical systems, like the formic acid dimer\cite{Fillaux2005, Miura2010, Marx2015, Ceriotti2016} and base pairs in DNA\cite{Pusuluk2018, Amico2008,  Chen2009}. 

In this Letter, we present an alternative approach that overcomes previous limitations and allows one to compute the R\'enyi entropy $S^\alpha_A$ for complex quantum systems, described by either model or \emph{ab initio} Hamiltonians. The latter is demonstrated by producing the first evidence of entanglement in the formic acid dimer. Our approach is based upon the combination of the path integral (PI) formalism and thermodynamic integration along appropriately defined paths.

\begin{figure}[t]
\includegraphics[width=1\linewidth]{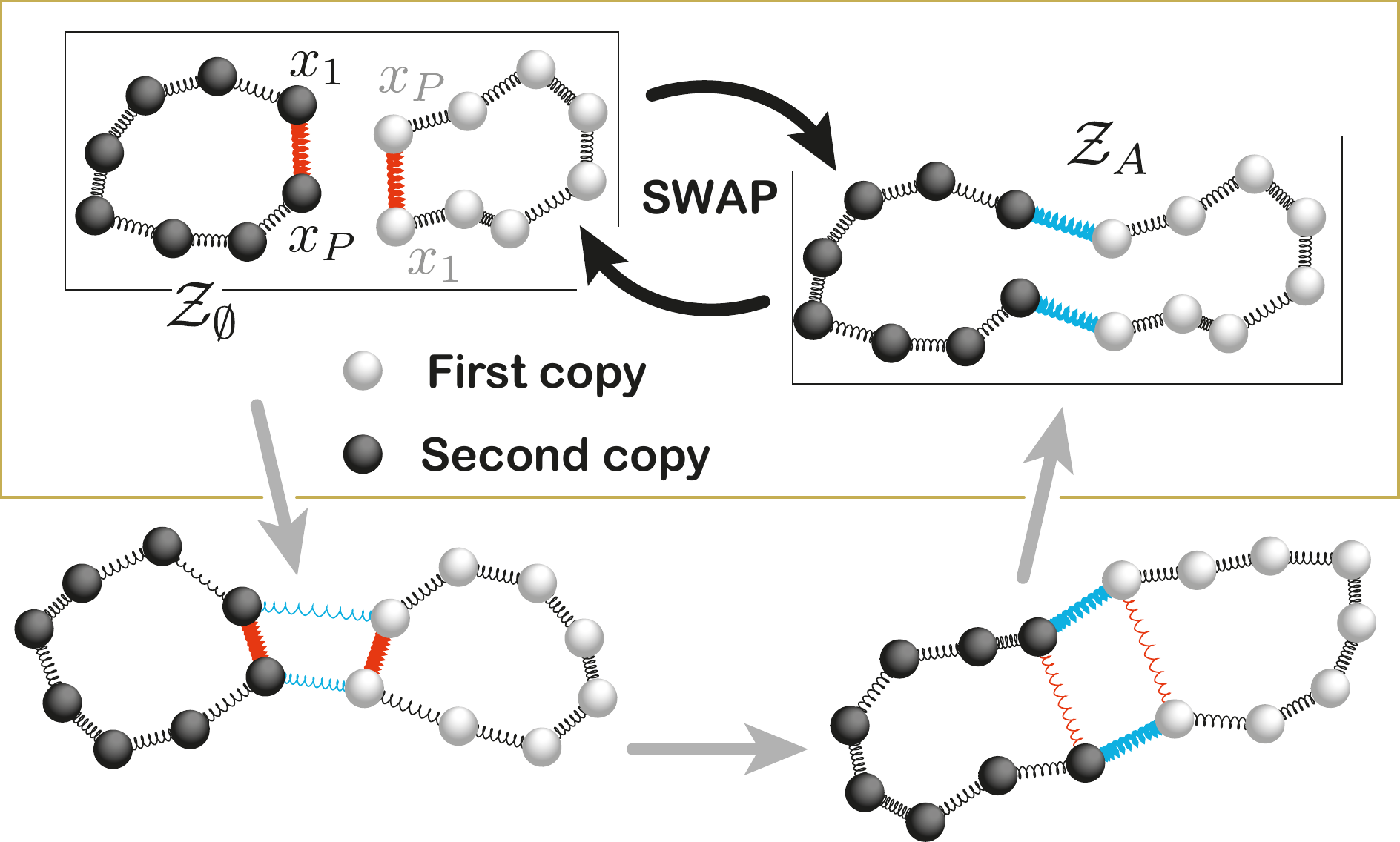}
\caption{\label{fig:ShematicOne} 
$\mathcal{Z}_{\emptyset}$ and $\mathcal{Z}_{A}$ for a quantum particle with $\alpha=2$ are shown in the upper boxes. $S^2$ can then be computed  by averaging the ``swap'' operator or by introducing intermediate steps (grey arrows) and averaging the spring interactions highlighted in colors (red, blue) over an appropriately defined path. The springs strength is proportional to color intensity.
}
\end{figure}

\emph{R\'enyi entropy with path integrals.}{---} 
In the PI formulations of statistical mechanics, the quantum density matrix $\rho$  at time $t$ is mapped to a classical counterpart by Wick rotating ($t\to-i\beta$) the quantum action and discretizing it into a classical Hamiltonian $H$\cite{Ceperley1995, Tuckerman}. In practical implementations, the imaginary time interval $[0,\beta=1/(k_B T)]$, $T$ being the temperature, is divided into a finite number of time steps, which are individual snapshots (or beads) that interact particle-wise in the imaginary time direction. One can then use importance sampling algorithms to evaluate the \miha{R\'enyi} entropy of $\alpha \in \mathbb{N} \setminus \{1\}$ by calculating the free energy difference between two statistical ensembles, namely $S_A^{\alpha} = \log(\mathcal{Z}_A / \mathcal{Z}_{\emptyset}) / (1 - \alpha)$\cite{Calabrese2004}. $\mathcal{Z}_\emptyset =[\text{Tr}\rho]^{\alpha}$ is the partition function of the $\emptyset$ ensemble, consisting of $\alpha$ independent copies of the system, and $\mathcal{Z}_A=\text{Tr}\rho_A^{\alpha}$ is the one for the  joint ensemble, where each particle belonging to the subsystem $A$ is replaced by one particle living in all the $\alpha$ copies. Thus, R\'enyi entropy depends \miha{only} on the free energy cost of changing the boundary conditions through the ``swap'' operator (see Fig.~\ref{fig:ShematicOne}) in the imaginary time direction.

\miha{This} free energy cost can be estimated by running a PI Monte Carlo simulation in one ensemble, say $\mathcal{Z}_{\emptyset}$, and averaging the exponent of the energy difference between the two boundary conditions at given $\beta$\cite{Hastings2010, Tony2010, FreeEnergyBook}, in order to evaluate
\begin{eqnarray}
\log\Big(\frac{\mathcal{Z}_{A}}{ \mathcal{Z}_{\emptyset}}\Big) = \log\Big\langle \exp(-\beta (H_A - H_{\emptyset}))\Big\rangle_{\mathcal{Z}_{\emptyset}}.
\label{eq:Pertrubation}
\end{eqnarray}
$H_{A, \emptyset}$ are classical Hamiltonians arising from the discretization of their corresponding quantum actions $\mathcal{Z}_{A,\emptyset}$. Though Eq.~\ref{eq:Pertrubation} is in principle exact, it implies that for increasing energy differences one has to wait exponentially longer times to sample high-energy configurations. A possible remedy for this slowing down is to split the calculation in shorter increments of smaller free-energy differences\cite{Hastings2010}, and to sample the ratio of the Metropolis–Hastings transition probabilities $\langle\text{min}(1, \exp(-\beta\Delta H))\rangle_{\mathcal{Z}_{\emptyset}} / \langle\text{min}(1, \exp(\beta \Delta H))\rangle_{\mathcal{Z}_{A}}$\cite{Bennett1976, Broecker2014}, rather than Eq.~\ref{eq:Pertrubation}. Following these procedures, reliable and groundbreaking entanglement entropy estimations were obtained in one-dimensional (1D) and two-dimensional (2D) spin chains\cite{Humeniuk2012, Luitz2014}, and cold atoms\cite{Herdman2014a}. Nevertheless, in larger or more strongly entangled systems, the consequently higher energy barriers can cause the aforementioned approach to fail.

\begin{figure*}[t]
\includegraphics[width=1\linewidth]{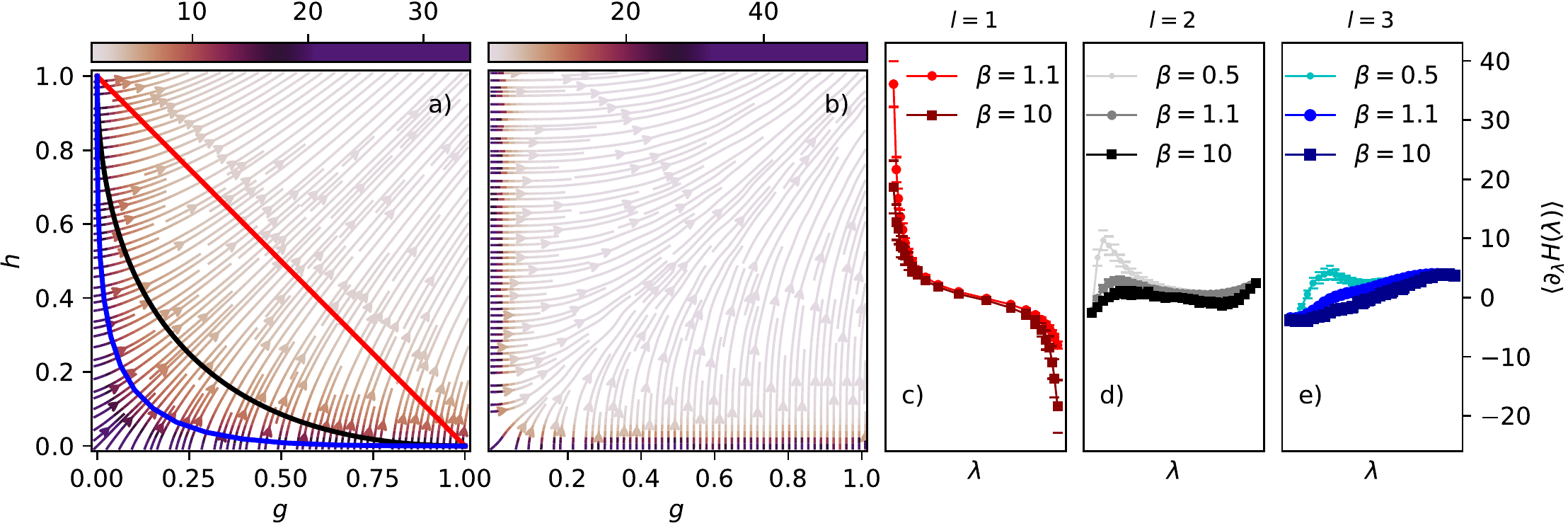}
\caption{\label{fig:IntegrationPath} 
a) Stream plot of the gradient field $(\langle K_{\emptyset} \rangle_{\mathcal{Z}[\lambda]}, \langle K_{A} \rangle_{\mathcal{Z}[\lambda]})$ for the single-particle 1D harmonic oscillator at $\beta=10$ as a function of $(g,h)$. Colors indicate the magnitude of the gradient, according to the palette above the frame. b) Stream plot of the variance field $(\text{var}[K_{\emptyset}], \text{var}[K_{A}])$. The black (blue) line represents the path in Eq. \ref{eq: IntegrationPathLambda} with $l=2$ ($l=3$). The red line is the linear path ($l=1$). Changing the path from linear to black or to blue, regularizes the integrand by cutting off its spikes at the edges, shown at different temperatures for $l=1,2,3$ in panels c), d) and e), respectively. 
}
\end{figure*}

\emph{Thermodynamic integration.}{---}
An alternative scheme is the thermodynamic integration based on a new partition function $\mathcal{Z}[\lambda]$, a differentiable function of the parameter $\lambda\in[0, 1]$, that connects the two ensembles\cite{Tony2010,Demidio2020}. Then, by using the relation:
\begin{eqnarray}
\log\Big(\frac{\mathcal{Z}_{A}}{ \mathcal{Z}_{\emptyset}}\Big) = -\beta\int_{0}^{1}\Big\langle \partial_{\lambda}H(\lambda)\Big\rangle_{\mathcal{Z}[\lambda]} d\lambda,
\label{eq:ThermodynamicIntegration}
\end{eqnarray}
and setting $\mathcal{Z}[0] = \mathcal{Z}_{\emptyset}$ and $\mathcal{Z}[1] = \mathcal{Z}_{A}$, one evaluates the entropy by sampling the derivative of the $H(\lambda)$ Hamiltonian appearing in $\mathcal{Z}[\lambda]$ and avoiding the exponential function present in Eq.~\ref{eq:Pertrubation}. Usually, the line integral in Eq.~\ref{eq:ThermodynamicIntegration} is performed numerically on a finite mesh. A common choice, previously used in the literature\cite{Buividovich2008}, is taking \emph{an integration path} that leads to $H(\lambda) = H_{\emptyset} + \lambda (H_A - H_{\emptyset})$. According to Eq.~\ref{eq:ThermodynamicIntegration}, then one has to average the energy difference $H_A - H_{\emptyset}$ over the $\mathcal{Z}[\lambda]$ ensemble. A representation of the $\lambda$-dependent part of $H(\lambda)$ is highlighted in colors in Fig.~\ref{fig:ShematicOne}. However, it was observed\cite{Buividovich2008} that the final value for the entropy is the result of an almost perfect cancellation between two possibly large contributions of opposite sign coming from $\lambda<1/2$ and $\lambda>1/2$, respectively. Moreover, at high temperature the integrand $\langle H_A - H_{\emptyset}\rangle_{\mathcal{Z}[\lambda]}$ diverges at the edges of the integration path (see Fig.~\ref{fig:IntegrationPath}c). Thus, to get reliable results high precision and a large number of integration steps are needed. The thermodynamic integration based on the simplest integration path is \miha{therefore} also  doomed to failure \miha{when faced with} large systems\cite{Humeniuk2012}.

In order to understand the origin of these drawbacks, \miha{we} focus on the harmonic oscillator and its \miha{R\'enyi} entropy of second order (i.e. with $\alpha=2$)\miha{, the exactly solvable minimal model, where the behaviour is reproduced}. In the PI formulation of $\mathcal{Z}_{\emptyset}$, each copy of the quantum harmonic oscillator of mass $m$ and frequency $\omega$ is described by a ring polymer with beads $\{x_k\}_{k=1,\ldots,P}$ connected by harmonic springs\cite{Tuckerman}, such that 
\begin{eqnarray}
H = \frac{m}{2}\sum_{k=1}^P \big[\frac{1}{\zeta^2\hbar^2P}(x_{k+1} - x_k)^2 + \frac{\omega^2}{P}x_k^2\big]\Biggr|_{x_{P+1} = x_1}.
\label{eq:HarmonicOscilator}
\end{eqnarray}
In the following, we will solve this model for $m=1$ and $\omega=1$. In Eq.~\ref{eq:HarmonicOscilator}, $\zeta=\beta/P$ is the imaginary time step which controls the discretization error and represents the inverse temperature of the classical analogue. Switching from the split ensemble to the joint one amounts to neglecting the harmonic interaction in $H_{\emptyset}$ between the beads $x_P$ and $x_1$ in both copies, and adding the interaction between $x_P$ of the first copy and $x_1$ of the second one, and its crossed term, in $H_A$ (see  Fig.~\ref{fig:ShematicOne}). The average energy of the joint interaction computed over $\mathcal{Z}_{\emptyset}$ is greater than the split one, because copies do not interact with each other in $\mathcal{Z}_{\emptyset}$ and, thus, they can be at relatively large distances. This contribution grows with the square of the inter-bead distance and with the phase-space size. Moreover, as the temperature increases, the interaction energy grows as $1 / \beta$ and de Broglie wavelength shrinks. Copies collapse to almost point-like particles and this further contributes to the diverging cost of joining them. This shows up in the large value of the integrand at $\lambda=0$ in the simplest integration scheme. \miha{Similarly, the variance of this contribution is large, since $\mathcal{Z}_{\emptyset}$ minimises the action without $H_A - H_{\emptyset}$.} \miha{An analogous }contribution but of opposite sign appears at $\lambda=1$, further increasing the overall uncertainty of the integral. Nevertheless, we found that if a more general path is considered, the pathological behavior of the integrand at the edges and its large variance can be avoided all together.

\emph{Path regularization.}{---}
In search of enhanced paths, we focus on the 2D parameter space $(g,h)$ of Hamiltonians $H(g, h)= H_{\emptyset} + (g - 1)K_{\emptyset} + h K_{A}$. The $K_{\emptyset,A}$ operators correspond to the \miha{interactions} that enforce the boundary conditions: $K_{\emptyset}$ $(K_{A})$ drives the intra (inter) -copy closure of the particle rings. These terms are depicted in red (blue) in Fig.~\ref{fig:ShematicOne}. Different paths connecting $(1,0)$ ($H=H_{\emptyset}$) to $(0,1)$ ($H=H_{A}$) can be compared by inspecting the gradient field of the energy $\boldsymbol{K} \equiv (\partial H/\partial g, \partial H/\partial h) = (\langle K_{\emptyset} \rangle_{\mathcal{Z}[\lambda]}, \langle K_{A} \rangle_{\mathcal{Z}[\lambda]})$, and the corresponding variance field $\text{var}[\boldsymbol{K}] \equiv (\text{var}[K_{\emptyset}], \text{var}[K_{A}])$ in the $(g,h)$ plane, shown in Figs.~\ref{fig:IntegrationPath}(a) and \ref{fig:IntegrationPath}(b), respectively. At each $(g,h)$ point, the direction of the field indicates the path $\boldsymbol{p} \equiv (g(\lambda),h(\lambda))$ that would yield the largest possible increment or the largest possible uncertainty to the line integral in Eq.~\ref{eq:ThermodynamicIntegration}, with magnitude represented by the color of the stream plot. For the 1D harmonic oscillator (Fig.~\ref{fig:IntegrationPath}c), we can clearly see that the path connecting linearly the two endpoints, i.e. the one that has commonly been employed so far, produces two spikes in $\langle \partial_{\lambda}H(\lambda)\rangle_{\mathcal{Z}[\lambda]}$ at $(1,0)$ and $(0,1)$, exactly where the scalar product of the path direction ($\boldsymbol{p^\prime} \equiv \partial_\lambda \boldsymbol{p}$) with the gradient field is the largest. This is understood once the integral in Eq.~\ref{eq:ThermodynamicIntegration} is recast in $-\beta \int_0^1 \boldsymbol{p'}\cdot\boldsymbol{K}  d\lambda$. The origin of spikes can be traced down to the $K_{A}$ growth along $(g, 0)$ and $K_{\emptyset}$ growth along $(0, h)$, as the temperature increases. \miha{These drawbacks are present in any system where intra-bead interaction is large.} In order to remove the spikes, one should therefore avoid moving towards directions where previously non-existing interactions are switched on in $H(g,h)$.

\begin{figure}[b]
\includegraphics[width=1\linewidth]{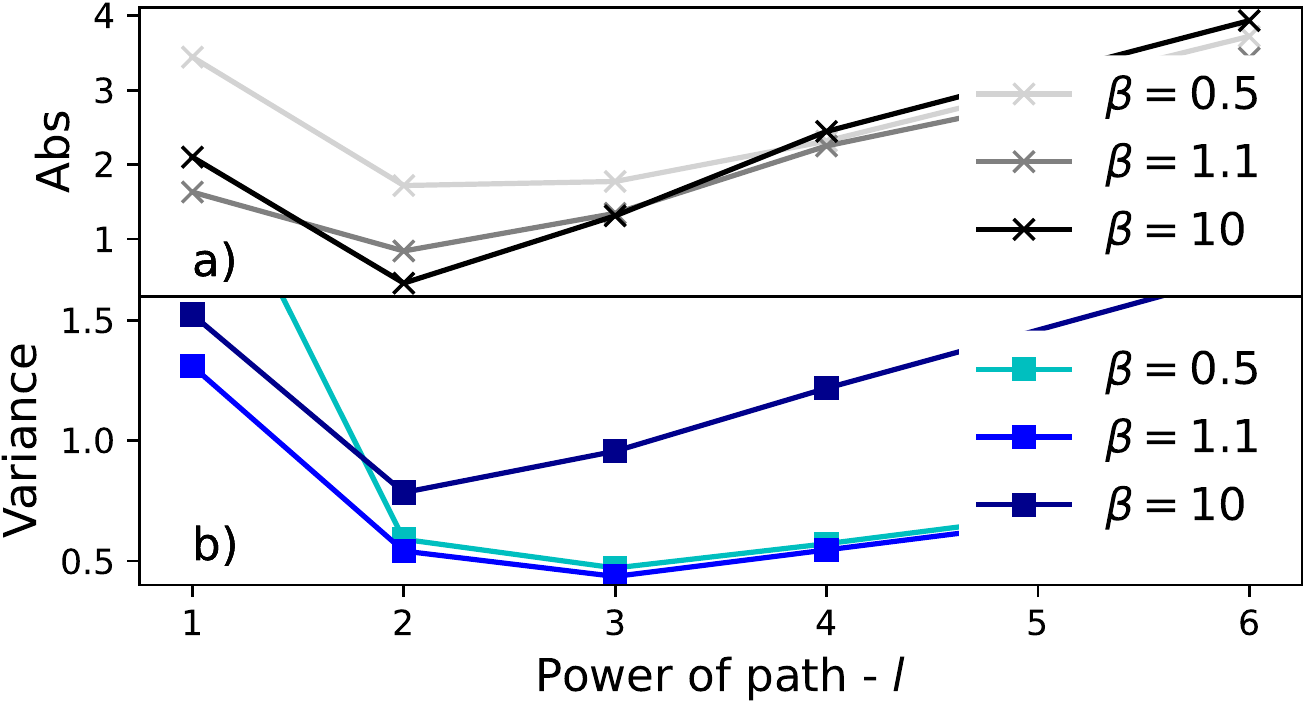}
\caption{\label{fig:Optimal_l}
\miha{Cost functionals\cite{SM} estimating a) excess area under the integrand ($F_\textrm{Abs}$ from Eq.~\ref{eq:functional1}) and b) variance of the integral ($F_\textrm{Variance}$ from Eq.~\ref{eq:functional2}) for different regularizing paths defined by the parameter $l$. Three values of the temperature are shown.}
}
\end{figure}

This analysis suggests that the optimal integration path is the one which minimizes 
\miha{\begin{equation}
F_\textrm{Abs}[\boldsymbol{p}] = \frac{1}{\int |\boldsymbol{p'}|d\lambda}\int_0^1 |\boldsymbol{p'}\cdot\boldsymbol{K}|d\lambda
\label{eq:functional1}
\end{equation}}
over the entire path. Another criterion can be derived by minimizing the full variance\cite{SM}, which implies the line minimization of
\miha{\begin{equation}
F_{Variance}[\boldsymbol{p}]=\int_0^1((\partial_\lambda g)^2,(\partial_\lambda h)^2) \cdot \text{var}[\boldsymbol{K}]d\lambda.
\label{eq:functional2}
\end{equation}}Either choice makes the integrand flat as a function of $\lambda$. This scheme acts therefore as a \emph{path regularization}. Not only the spikes at the endpoints are cut off, but the line integral can also be computed on a much coarser grid, speeding up the calculation and reducing both deterministic and stochastic errors.

It is clear that the full path optimization would not be affordable in the most general case. However, the simple 1D quantum oscillator, where the path search can be carried out systematically, provides a shape that is transferable to complex quantum many-body systems. $\boldsymbol{p}$ can then be parametrized as differentiable curve:
\begin{equation}
\big(g(\lambda),h(\lambda)\big)=\big((1 - \lambda)^l,\lambda^l\big) ~~~~\textrm{with $\lambda \in [0,1]$}.
\label{eq: IntegrationPathLambda}
\end{equation}
If rescaled by the proton mass, the reported behavior of the 1D harmonic oscillator spans a physically relevant range of parameters with temperatures going from 4.3~K ($\beta=40$) to 344~K ($\beta=0.5$) and with a vibrational frequency of 5122 cm$^{-1}$. In this realistic regime, it turns out that $l=2$ optimizes the path based on $|\boldsymbol{p^\prime} \cdot \boldsymbol{K}|$ \miha{(Fig.~\ref{fig:Optimal_l}(a))}, while $l=3$ is the optimal power law based on $\text{var}[\boldsymbol{K}]$ \miha{(Fig.~\ref{fig:Optimal_l}(b))}. Nevertheless, the latter is the best choice in \emph{ab initio} systems over a large range of temperatures\cite{SM}. \michele{Indeed, these systems share a similar intra-bead interaction to the one of the harmonic oscillator studied above, provided by the quantum kinetic term of the \emph{ab initio} action.}

\begin{figure}[t]
\includegraphics[width=1\linewidth]{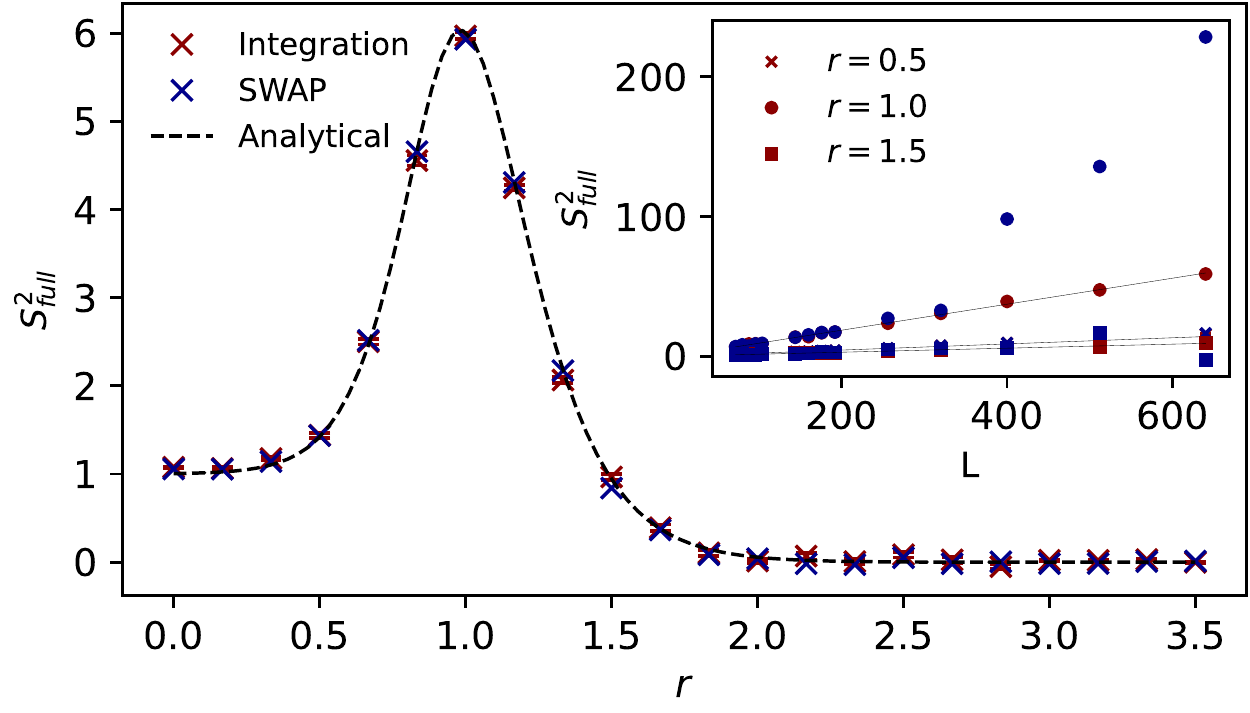}
\caption{\label{fig:IsingModel} Ising model. $r$ dependence of the full R\'enyi entropy of second order ($S^2_\textrm{full} / \log(2)$) computed with thermodynamic integration via path regularization (red crosses) and transition probability sampling based on the swap operator (blue crosses) compared with \miha{analytical} results (dashed-black line) for $L=64$ and $\beta = 3$. \textit{Inset}: Comparison of system-size scaling for both methods at different $r$. 
}
\end{figure}

\emph{1D Ising model in transverse magnetic field.}{---}
The harmonic oscillator is however a too simple model to possess any entanglement. In a system with a larger number of interacting particles, the coefficients $(g,h)$ change the boundary condition of the full subsystem. To explicitly check the generality of the \emph{path regularization} approach, we now study the 1D Ising model in transverse magnetic field with periodic boundary conditions. Its Hamiltonian reads
\begin{eqnarray}
\begin{matrix}
H = \sum_i\sigma_i^z\sigma_{i+1}^z + r\sigma_i^x,
\end{matrix}
\label{eq: IsingModel}
\end{eqnarray}
where $\sigma_i^{x,z}$ are Pauli matrices acting on $i$-th site, and $r$ the strength of the magnetic field in the $x$ direction. Due to its integrability \miha{(it can be analytically solved using Jordan-Wigner transformation}\cite{Mbeng2020}\miha{)}, it represents an ideal benchmark for our path regularization in an extended system. It undergoes a quantum phase transition at $r=1$.

Within the PI framework, Eq.~\ref{eq: IsingModel} is mapped into the 2D classical anisotropic Ising model\cite{SM}. The interaction in the imaginary time direction of the classical counterpart diverges as $r\to 0$, causing effects analogous to the temperature increase in the harmonic oscillator. At small $r$, the path in Eq.~\ref{eq: IntegrationPathLambda} successfully kills large fluctuations appearing at the endpoints if \miha{$l=3$} is used. In this case, there is still room for improving the thermodynamic integration path. \miha{Indeed, in this model system, the magnitude of the endpoints still grows with the interaction strength. This can be cured by rescaling the $\lambda$ parameter\cite{SM}, which provides an additional freedom to optimize the path. Rescaling $\lambda$ leaves the shape of the path unchanged but it varies the ``speed'' (i.e. the integration points density) along the thermodynamic trajectory.} To compare our method with the sampling algorithms based on the swap operator (Eq.~\ref{eq:Pertrubation}), like the one of Ref.~\onlinecite{Humeniuk2012}, we computed\cite{Srdinsek2021github} \miha{the R\'enyi entropy of second order $S^2_\textrm{full}$ with the whole system as a subsystem}. It is the hardest quantity to evaluate. The full entropies obtained with the swap scheme and the \emph{path regularization} both agree well with analytical results over a wide range of magnetic field strengths (Fig.~\ref{fig:IsingModel}). $S^2_\textrm{full}$ nicely captures the quantum phase transition at $r=1$, by showing a clear peak. In order to estimate the maximum system size that the \emph{path regularization} procedure can afford, we pushed the entropy calculation to very long spin chains, where the thermodynamic limit is reached. In the limit of large $L$, the full entropy scales as $L$. From the inset of Fig.~\ref{fig:IsingModel}, it can be seen that this limit is reached at relatively small system sizes. It is also seen that our procedure outperforms the swap-based one, since the former can still be applied to systems larger than 600 sites where $S^2_\textrm{full}$ exceeds the value of 50, while the latter is broken already before 400 sites\cite{Humeniuk2012}. Indeed, one of the strengths of the path regularization method is that the number of integration steps remains constant with the subsystem size. By increasing the level of entanglement, the time cost grows linearly, while in methods based on Eq.~\ref{eq:Pertrubation}, it grows exponentially.
\begin{figure*}[t]
\includegraphics[width=0.9\linewidth]{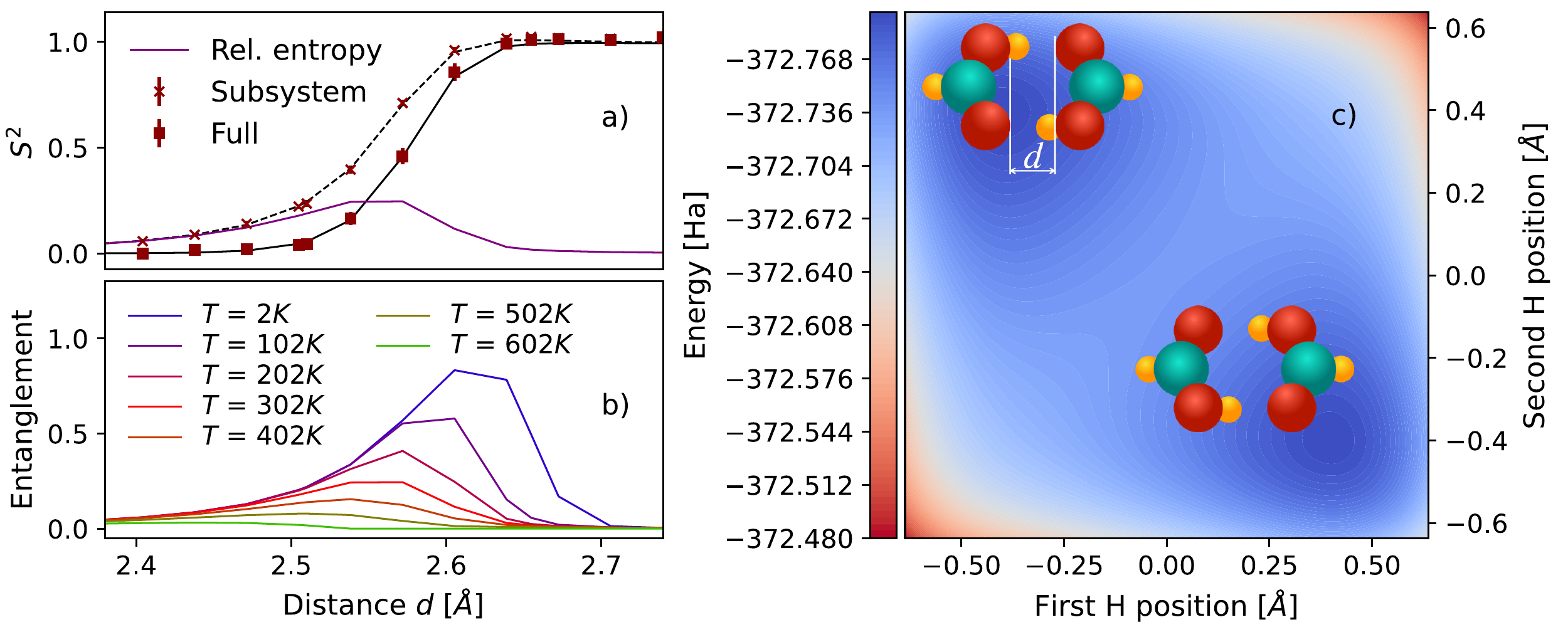}
\caption{\label{fig:DoubleWell} a) Full (black solid line) \hig{entropy and the entropy of a single proton subspace ($S^2_{A} / \log(2)$) (black dashed line) as a function of the distance $d$} between the formic acid molecules at $300K$. The difference between the two (``relative entropy'') \miha{-- the lower bound of entanglement --} is also shown (purple solid line). b) Temperature dependence of relative entropy. c) 2D PES spanned by the position of the hydrogen atoms along the hydrogen bonds at $d = 2.87${\AA}.
}
\end{figure*}

\emph{A realistic system: the formic acid dimer.}{---} 
By means of the \emph{path regularization} scheme, the calculation of the R\'enyi entropy for realistic systems becomes feasible. As an illustrative example, we take here the formic acid dimer, a minimal model of biochemical physics. This is a system of two molecules that form a dimer via a double hydrogen bond (Fig.~\ref{fig:DoubleWell}c) Due to the 180$^\circ$ rotation symmetry around the axis connecting the carbon atoms, the potential energy surface (PES) has two minima, which correspond to the two hydrogen configurations shown in Fig.~\ref{fig:DoubleWell}. They are separated by a barrier that grows with the inter-dimer distance $d$ between the oxygen atoms ($d = 2.7${\AA} in equilibrium\cite{Miura2010, Tachikawa2020}), which determines also the hydrogen bond stretch. During the double proton transfer the two molecules get closer, up to $d=2.4${\AA}, and the barrier dwindles. Due to the light hydrogen mass, at intermediate distances quantum effect become prominent\cite{Fillaux2005, Miura2010, Marx2015, Ceriotti2016} and the barrier is low enough that the two configurations are expected to be entangled, leading to a proton concerted motion.

In our simulations, we restricted the protons to move along the hydrogen bond and we evaluated the PES as a function of $d$ by means of the coupled cluster single double perturbative triple (CCSD(T)) method\cite{SM}. At each distance, we ran a PI Monte Carlo simulation of the two protons in a projected PES with our thermodynamic integration scheme\cite{Srdinsek2021github}. As the potential is simple enough, the outcome of these simulations can be compared against results obtained with the exact diagonalization in a discretized space. This comparison further assesses the robustness of the \emph{path regularization} in Eq.\ref{eq: IntegrationPathLambda}, with the optimal power of $l=3$ for \emph{ab initio} systems. Since the R\'enyi entropy of the 2-proton subspace (``full'' entropy) is nonzero at distances where the system is in a mixed state, we can extract a lower bound for quantum correlations\cite{Wilde2011} by computing the difference between the full and the \hig{R\'enyi entropy of the single proton subspace (entanglement entropy if in a pure state)}. The results show that the entanglement is present along the full range of distances, explored during the double proton transfer ($d = 2.4-2.7${\AA}\cite{Miura2010}). However, the temperature plays a major role. At room temperature the entanglement is considerably lower compared to the one at $100K$. Remarkably, it persists up to temperatures as large as $500K$ for some intermolecular distance. We note that at high temperatures the thermal motion of the full molecular complex must be taken into account. Nevertheless, the effect is so strong that the entanglement should still be relevant well above room temperature even in this case. 

\emph{Conclusions.}{---}
In this Letter, we introduced a path regularization scheme that allows an efficient and stable calculation of the \miha{Quantum} R\'enyi entropy via a thermodynamic integration performed within the path integral framework. The proposed regularization defines an optimal thermodynamic path that smoothly changes the imaginary time boundary conditions of the quantum partition function, by avoiding slowly convergent contributions and by yielding a low-variance estimate of the \miha{R\'enyi} entropy. The method has been shown to be efficient in the 1D Ising model with transverse magnetic field, where we reached very large subsystem sizes. It also allowed us to perform a first \textit{ab initio} evaluation of the R\'enyi entropy for the concerted hydrogen motion in the formic acid dimer. The path regularization makes the evaluation of the R\'enyi entropy feasible for model and real systems, comprising large sizes and/or complex interactions, for which the R\'enyi entropy analysis was previously inaccessible.

\begin{acknowledgments}

We thank the support of the HPCaVe computational platform of Sorbonne University, where the main calculations have been performed. We are grateful for the environment provided by ISCD and its MAESTRO junior team. This  work  was partially supported  by  the  European  Centre of Excellence in Exascale Computing TREX-Targeting Real Chemical Accuracy at the Exascale, funded by the European Union’s Horizon 2020 Research and Innovation program under Grant Agreement No.~952165.

\end{acknowledgments}

\bibliography{PIRenyiRegPaths}

\begin{thebibliography}{61}%
\makeatletter
\providecommand \@ifxundefined [1]{%
 \@ifx{#1\undefined}
}%
\providecommand \@ifnum [1]{%
 \ifnum #1\expandafter \@firstoftwo
 \else \expandafter \@secondoftwo
 \fi
}%
\providecommand \@ifx [1]{%
 \ifx #1\expandafter \@firstoftwo
 \else \expandafter \@secondoftwo
 \fi
}%
\providecommand \natexlab [1]{#1}%
\providecommand \enquote  [1]{``#1''}%
\providecommand \bibnamefont  [1]{#1}%
\providecommand \bibfnamefont [1]{#1}%
\providecommand \citenamefont [1]{#1}%
\providecommand \href@noop [0]{\@secondoftwo}%
\providecommand \href [0]{\begingroup \@sanitize@url \@href}%
\providecommand \@href[1]{\@@startlink{#1}\@@href}%
\providecommand \@@href[1]{\endgroup#1\@@endlink}%
\providecommand \@sanitize@url [0]{\catcode `\\12\catcode `\$12\catcode
  `\&12\catcode `\#12\catcode `\^12\catcode `\_12\catcode `\%12\relax}%
\providecommand \@@startlink[1]{}%
\providecommand \@@endlink[0]{}%
\providecommand \url  [0]{\begingroup\@sanitize@url \@url }%
\providecommand \@url [1]{\endgroup\@href {#1}{\urlprefix }}%
\providecommand \urlprefix  [0]{URL }%
\providecommand \Eprint [0]{\href }%
\providecommand \doibase [0]{https://doi.org/}%
\providecommand \selectlanguage [0]{\@gobble}%
\providecommand \bibinfo  [0]{\@secondoftwo}%
\providecommand \bibfield  [0]{\@secondoftwo}%
\providecommand \translation [1]{[#1]}%
\providecommand \BibitemOpen [0]{}%
\providecommand \bibitemStop [0]{}%
\providecommand \bibitemNoStop [0]{.\EOS\space}%
\providecommand \EOS [0]{\spacefactor3000\relax}%
\providecommand \BibitemShut  [1]{\csname bibitem#1\endcsname}%
\let\auto@bib@innerbib\@empty
\bibitem [{\citenamefont {Horodecki}\ \emph {et~al.}(2009)\citenamefont
  {Horodecki}, \citenamefont {Horodecki}, \citenamefont {Horodecki},\ and\
  \citenamefont {Horodecki}}]{Horodecki2009}%
  \BibitemOpen
  \bibfield  {author} {\bibinfo {author} {\bibfnamefont {R.}~\bibnamefont
  {Horodecki}}, \bibinfo {author} {\bibfnamefont {P.}~\bibnamefont
  {Horodecki}}, \bibinfo {author} {\bibfnamefont {M.}~\bibnamefont
  {Horodecki}},\ and\ \bibinfo {author} {\bibfnamefont {K.}~\bibnamefont
  {Horodecki}},\ }\bibfield  {title} {\bibinfo {title} {{Quantum
  entanglement}},\ }\href {https://doi.org/10.1103/RevModPhys.81.865}
  {\bibfield  {journal} {\bibinfo  {journal} {Rev. Mod. Phys.}\ }\textbf
  {\bibinfo {volume} {81}},\ \bibinfo {pages} {865} (\bibinfo {year}
  {2009})}\BibitemShut {NoStop}%
\bibitem [{\citenamefont {Hastings}\ \emph {et~al.}(2010)\citenamefont
  {Hastings}, \citenamefont {Gonz{\'{a}}lez}, \citenamefont {Kallin},\ and\
  \citenamefont {Melko}}]{Hastings2010}%
  \BibitemOpen
  \bibfield  {author} {\bibinfo {author} {\bibfnamefont {M.~B.}\ \bibnamefont
  {Hastings}}, \bibinfo {author} {\bibfnamefont {I.}~\bibnamefont
  {Gonz{\'{a}}lez}}, \bibinfo {author} {\bibfnamefont {A.~B.}\ \bibnamefont
  {Kallin}},\ and\ \bibinfo {author} {\bibfnamefont {R.~G.}\ \bibnamefont
  {Melko}},\ }\bibfield  {title} {\bibinfo {title} {{Measuring renyi
  entanglement entropy in quantum Monte Carlo simulations}},\ }\href
  {https://doi.org/10.1103/PhysRevLett.104.157201} {\bibfield  {journal}
  {\bibinfo  {journal} {Phys. Rev. Lett.}\ }\textbf {\bibinfo {volume} {104}},\
  \bibinfo {pages} {157201} (\bibinfo {year} {2010})}\BibitemShut {NoStop}%
\bibitem [{\citenamefont {Humeniuk}\ and\ \citenamefont
  {Roscilde}(2012)}]{Humeniuk2012}%
  \BibitemOpen
  \bibfield  {author} {\bibinfo {author} {\bibfnamefont {S.}~\bibnamefont
  {Humeniuk}}\ and\ \bibinfo {author} {\bibfnamefont {T.}~\bibnamefont
  {Roscilde}},\ }\bibfield  {title} {\bibinfo {title} {{Quantum Monte Carlo
  calculation of entanglement R{\'{e}}nyi entropies for generic quantum
  systems}},\ }\href {https://doi.org/10.1103/PhysRevB.86.235116} {\bibfield
  {journal} {\bibinfo  {journal} {Phys. Rev. B}\ }\textbf {\bibinfo {volume}
  {86}},\ \bibinfo {pages} {235116} (\bibinfo {year} {2012})}\BibitemShut
  {NoStop}%
\bibitem [{\citenamefont {Alba}(2017)}]{Alba2017}%
  \BibitemOpen
  \bibfield  {author} {\bibinfo {author} {\bibfnamefont {V.}~\bibnamefont
  {Alba}},\ }\bibfield  {title} {\bibinfo {title} {{Out-of-equilibrium protocol
  for Renyi entropies via the Jarzynski equality}},\ }\href
  {https://doi.org/10.1103/PhysRevE.95.062132} {\bibfield  {journal} {\bibinfo
  {journal} {Phys. Rev. E}\ }\textbf {\bibinfo {volume} {95}},\ \bibinfo
  {pages} {062132} (\bibinfo {year} {2017})}\BibitemShut {NoStop}%
\bibitem [{\citenamefont {DEmidio}(2020)}]{Demidio2020}%
  \BibitemOpen
  \bibfield  {author} {\bibinfo {author} {\bibfnamefont {J.}~\bibnamefont
  {DEmidio}},\ }\bibfield  {title} {\bibinfo {title} {{Entanglement Entropy
  from Nonequilibrium Work}},\ }\href
  {https://doi.org/10.1103/PhysRevLett.124.110602} {\bibfield  {journal}
  {\bibinfo  {journal} {Phys. Rev. Lett.}\ }\textbf {\bibinfo {volume} {124}},\
  \bibinfo {pages} {110602} (\bibinfo {year} {2020})}\BibitemShut {NoStop}%
\bibitem [{\citenamefont {White}\ and\ \citenamefont
  {Meirovitch}(2004)}]{White2004}%
  \BibitemOpen
  \bibfield  {author} {\bibinfo {author} {\bibfnamefont {R.~P.}\ \bibnamefont
  {White}}\ and\ \bibinfo {author} {\bibfnamefont {H.}~\bibnamefont
  {Meirovitch}},\ }\bibfield  {title} {\bibinfo {title} {{A simulation method
  for calculating the absolute entropy and free energy of fluids: Application
  to liquid argon and water}},\ }\href
  {https://doi.org/10.1073/pnas.0308197101} {\bibfield  {journal} {\bibinfo
  {journal} {Proc. Natl. Acad. Sci. U. S. A.}\ }\textbf {\bibinfo {volume}
  {101}},\ \bibinfo {pages} {9235} (\bibinfo {year} {2004})}\BibitemShut
  {NoStop}%
\bibitem [{\citenamefont {Do}\ and\ \citenamefont {Wheatley}(2013)}]{Do2013}%
  \BibitemOpen
  \bibfield  {author} {\bibinfo {author} {\bibfnamefont {H.}~\bibnamefont
  {Do}}\ and\ \bibinfo {author} {\bibfnamefont {R.~J.}\ \bibnamefont
  {Wheatley}},\ }\bibfield  {title} {\bibinfo {title} {{Density of states
  partitioning method for calculating the free energy of solids}},\ }\href
  {https://doi.org/10.1021/ct3007056} {\bibfield  {journal} {\bibinfo
  {journal} {J. Chem. Theory Comput.}\ }\textbf {\bibinfo {volume} {9}},\
  \bibinfo {pages} {165} (\bibinfo {year} {2013})}\BibitemShut {NoStop}%
\bibitem [{\citenamefont {Lelievre}\ \emph {et~al.}(2010)\citenamefont
  {Lelievre}, \citenamefont {Rousset},\ and\ \citenamefont
  {Stoltz}}]{Tony2010}%
  \BibitemOpen
  \bibfield  {author} {\bibinfo {author} {\bibfnamefont {T.}~\bibnamefont
  {Lelievre}}, \bibinfo {author} {\bibfnamefont {M.}~\bibnamefont {Rousset}},\
  and\ \bibinfo {author} {\bibfnamefont {G.}~\bibnamefont {Stoltz}},\ }\href
  {https://doi.org/10.1142/p579} {\emph {\bibinfo {title} {{Free Energy
  Computations}}}}\ (\bibinfo  {publisher} {Imperial College Press},\ \bibinfo
  {year} {2010})\BibitemShut {NoStop}%
\bibitem [{\citenamefont {Luitz}\ \emph {et~al.}(2014)\citenamefont {Luitz},
  \citenamefont {Plat}, \citenamefont {Laflorencie},\ and\ \citenamefont
  {Alet}}]{Luitz2014}%
  \BibitemOpen
  \bibfield  {author} {\bibinfo {author} {\bibfnamefont {D.~J.}\ \bibnamefont
  {Luitz}}, \bibinfo {author} {\bibfnamefont {X.}~\bibnamefont {Plat}},
  \bibinfo {author} {\bibfnamefont {N.}~\bibnamefont {Laflorencie}},\ and\
  \bibinfo {author} {\bibfnamefont {F.}~\bibnamefont {Alet}},\ }\bibfield
  {title} {\bibinfo {title} {{Improving entanglement and thermodynamic
  R{\'{e}}nyi entropy measurements in quantum Monte Carlo}},\ }\href
  {https://doi.org/10.1103/PhysRevB.90.125105} {\bibfield  {journal} {\bibinfo
  {journal} {Phys. Rev. B}\ }\textbf {\bibinfo {volume} {90}},\ \bibinfo
  {pages} {125105} (\bibinfo {year} {2014})}\BibitemShut {NoStop}%
\bibitem [{\citenamefont {Preskill}(2016)}]{Preskill2016}%
  \BibitemOpen
  \bibfield  {author} {\bibinfo {author} {\bibfnamefont {J.}~\bibnamefont
  {Preskill}},\ }\href {http://arxiv.org/abs/1604.07450} {\emph {\bibinfo
  {title} {{Quantum Information Chapter 10. Quantum Shannon Theory}}}},\
  \bibinfo {number} {April}\ (\bibinfo  {publisher} {arXiv:1604.07450},\
  \bibinfo {year} {2016})\BibitemShut {NoStop}%
\bibitem [{\citenamefont {Cover}\ and\ \citenamefont
  {Thomas}(2005)}]{Cover1938}%
  \BibitemOpen
  \bibfield  {author} {\bibinfo {author} {\bibfnamefont {T.~M.}\ \bibnamefont
  {Cover}}\ and\ \bibinfo {author} {\bibfnamefont {J.~A.}\ \bibnamefont
  {Thomas}},\ }\href {https://doi.org/10.1002/047174882X} {\emph {\bibinfo
  {title} {Elem. Inf. Theory}}},\ \bibinfo {edition} {2nd}\ ed.\ (\bibinfo
  {publisher} {John Wiley and Sons, Inc., Hoboken, New Jersey.},\ \bibinfo
  {year} {2005})\BibitemShut {NoStop}%
\bibitem [{\citenamefont {Horodecki}\ and\ \citenamefont
  {Horodecki}(1996)}]{Horodecki1996}%
  \BibitemOpen
  \bibfield  {author} {\bibinfo {author} {\bibfnamefont {R.}~\bibnamefont
  {Horodecki}}\ and\ \bibinfo {author} {\bibfnamefont {M.}~\bibnamefont
  {Horodecki}},\ }\bibfield  {title} {\bibinfo {title} {{Information-theoretic
  aspects of inseparability of mixed states}},\ }\bibfield  {journal} {\bibinfo
   {journal} {Phys. Rev. A}\ }\textbf {\bibinfo {volume} {54}},\ \href
  {https://doi.org/10.1103/PhysRevA.54.1838} {10.1103/PhysRevA.54.1838}
  (\bibinfo {year} {1996})\BibitemShut {NoStop}%
\bibitem [{\citenamefont {Vidal}\ \emph {et~al.}(2003)\citenamefont {Vidal},
  \citenamefont {Latorre}, \citenamefont {Rico},\ and\ \citenamefont
  {Kitaev}}]{Vidal2003}%
  \BibitemOpen
  \bibfield  {author} {\bibinfo {author} {\bibfnamefont {G.}~\bibnamefont
  {Vidal}}, \bibinfo {author} {\bibfnamefont {J.~I.}\ \bibnamefont {Latorre}},
  \bibinfo {author} {\bibfnamefont {E.}~\bibnamefont {Rico}},\ and\ \bibinfo
  {author} {\bibfnamefont {A.}~\bibnamefont {Kitaev}},\ }\bibfield  {title}
  {\bibinfo {title} {{Entanglement in Quantum Critical Phenomena}},\ }\href
  {https://doi.org/10.1103/PhysRevLett.90.227902} {\bibfield  {journal}
  {\bibinfo  {journal} {Phys. Rev. Lett.}\ }\textbf {\bibinfo {volume} {90}},\
  \bibinfo {pages} {227902} (\bibinfo {year} {2003})}\BibitemShut {NoStop}%
\bibitem [{\citenamefont {Calabrese}\ and\ \citenamefont
  {Cardy}(2004)}]{Calabrese2004}%
  \BibitemOpen
  \bibfield  {author} {\bibinfo {author} {\bibfnamefont {P.}~\bibnamefont
  {Calabrese}}\ and\ \bibinfo {author} {\bibfnamefont {J.}~\bibnamefont
  {Cardy}},\ }\bibfield  {title} {\bibinfo {title} {Entanglement entropy and
  quantum field theory},\ }\href
  {https://doi.org/10.1088/1742-5468/2004/06/p06002} {\bibfield  {journal}
  {\bibinfo  {journal} {J. Stat. Mech. Theory Exp.}\ }\textbf {\bibinfo
  {volume} {2004}},\ \bibinfo {pages} {06002} (\bibinfo {year}
  {2004})}\BibitemShut {NoStop}%
\bibitem [{\citenamefont {Wolf}\ \emph {et~al.}(2008)\citenamefont {Wolf},
  \citenamefont {Verstraete}, \citenamefont {Hastings},\ and\ \citenamefont
  {Cirac}}]{Cirac2008}%
  \BibitemOpen
  \bibfield  {author} {\bibinfo {author} {\bibfnamefont {M.~M.}\ \bibnamefont
  {Wolf}}, \bibinfo {author} {\bibfnamefont {F.}~\bibnamefont {Verstraete}},
  \bibinfo {author} {\bibfnamefont {M.~B.}\ \bibnamefont {Hastings}},\ and\
  \bibinfo {author} {\bibfnamefont {J.~I.}\ \bibnamefont {Cirac}},\ }\bibfield
  {title} {\bibinfo {title} {{Area laws in quantum systems: Mutual information
  and correlations}},\ }\href {https://doi.org/10.1103/PhysRevLett.100.070502}
  {\bibfield  {journal} {\bibinfo  {journal} {Phys. Rev. Lett.}\ }\textbf
  {\bibinfo {volume} {100}},\ \bibinfo {pages} {070502} (\bibinfo {year}
  {2008})}\BibitemShut {NoStop}%
\bibitem [{\citenamefont {Flammia}\ \emph {et~al.}(2009)\citenamefont
  {Flammia}, \citenamefont {Hamma}, \citenamefont {Hughes},\ and\ \citenamefont
  {Wen}}]{Flammia2009}%
  \BibitemOpen
  \bibfield  {author} {\bibinfo {author} {\bibfnamefont {S.~T.}\ \bibnamefont
  {Flammia}}, \bibinfo {author} {\bibfnamefont {A.}~\bibnamefont {Hamma}},
  \bibinfo {author} {\bibfnamefont {T.~L.}\ \bibnamefont {Hughes}},\ and\
  \bibinfo {author} {\bibfnamefont {X.~G.}\ \bibnamefont {Wen}},\ }\bibfield
  {title} {\bibinfo {title} {{Topological entanglement R{\'{e}}nyi entropy and
  reduced density matrix structure}},\ }\href
  {https://doi.org/10.1103/PhysRevLett.103.261601} {\bibfield  {journal}
  {\bibinfo  {journal} {Phys. Rev. Lett.}\ }\textbf {\bibinfo {volume} {103}},\
  \bibinfo {pages} {261601} (\bibinfo {year} {2009})}\BibitemShut {NoStop}%
\bibitem [{\citenamefont {Metlitski}\ \emph {et~al.}(2009)\citenamefont
  {Metlitski}, \citenamefont {Fuertes},\ and\ \citenamefont
  {Sachdev}}]{Metlitski2009}%
  \BibitemOpen
  \bibfield  {author} {\bibinfo {author} {\bibfnamefont {M.~A.}\ \bibnamefont
  {Metlitski}}, \bibinfo {author} {\bibfnamefont {C.~A.}\ \bibnamefont
  {Fuertes}},\ and\ \bibinfo {author} {\bibfnamefont {S.}~\bibnamefont
  {Sachdev}},\ }\bibfield  {title} {\bibinfo {title} {{Entanglement entropy in
  the O(N) model}},\ }\href {https://doi.org/10.1103/PhysRevB.80.115122}
  {\bibfield  {journal} {\bibinfo  {journal} {Phys. Rev. B}\ }\textbf {\bibinfo
  {volume} {80}},\ \bibinfo {pages} {115122} (\bibinfo {year}
  {2009})}\BibitemShut {NoStop}%
\bibitem [{\citenamefont {Singh}\ \emph {et~al.}(2011)\citenamefont {Singh},
  \citenamefont {Hastings}, \citenamefont {Kallin},\ and\ \citenamefont
  {Melko}}]{Singh2011}%
  \BibitemOpen
  \bibfield  {author} {\bibinfo {author} {\bibfnamefont {R.~R.~P.}\
  \bibnamefont {Singh}}, \bibinfo {author} {\bibfnamefont {M.~B.}\ \bibnamefont
  {Hastings}}, \bibinfo {author} {\bibfnamefont {A.~B.}\ \bibnamefont
  {Kallin}},\ and\ \bibinfo {author} {\bibfnamefont {R.~G.}\ \bibnamefont
  {Melko}},\ }\bibfield  {title} {\bibinfo {title} {{Finite-temperature
  critical behavior of mutual information}},\ }\href
  {https://doi.org/10.1103/PhysRevLett.106.135701} {\bibfield  {journal}
  {\bibinfo  {journal} {Phys. Rev. Lett.}\ }\textbf {\bibinfo {volume} {106}},\
  \bibinfo {pages} {135701} (\bibinfo {year} {2011})}\BibitemShut {NoStop}%
\bibitem [{\citenamefont {Romera}\ and\ \citenamefont
  {Nagy}(2011)}]{Romera2011}%
  \BibitemOpen
  \bibfield  {author} {\bibinfo {author} {\bibfnamefont {E.}~\bibnamefont
  {Romera}}\ and\ \bibinfo {author} {\bibfnamefont {{\'{A}}.}~\bibnamefont
  {Nagy}},\ }\bibfield  {title} {\bibinfo {title} {{R{\'{e}}nyi entropy and
  quantum phase transition in the Dicke model}},\ }\href
  {https://doi.org/10.1016/j.physleta.2011.06.046} {\bibfield  {journal}
  {\bibinfo  {journal} {Phys. Lett. A}\ }\textbf {\bibinfo {volume} {375}},\
  \bibinfo {pages} {3066} (\bibinfo {year} {2011})}\BibitemShut {NoStop}%
\bibitem [{\citenamefont {Herdman}\ \emph
  {et~al.}(2014{\natexlab{a}})\citenamefont {Herdman}, \citenamefont {Inglis},
  \citenamefont {Roy}, \citenamefont {Melko},\ and\ \citenamefont {{Del
  Maestro}}}]{Herdman2014}%
  \BibitemOpen
  \bibfield  {author} {\bibinfo {author} {\bibfnamefont {C.~M.}\ \bibnamefont
  {Herdman}}, \bibinfo {author} {\bibfnamefont {S.}~\bibnamefont {Inglis}},
  \bibinfo {author} {\bibfnamefont {P.~N.}\ \bibnamefont {Roy}}, \bibinfo
  {author} {\bibfnamefont {R.~G.}\ \bibnamefont {Melko}},\ and\ \bibinfo
  {author} {\bibfnamefont {A.}~\bibnamefont {{Del Maestro}}},\ }\bibfield
  {title} {\bibinfo {title} {{Path-integral Monte Carlo method for R{\'{e}}nyi
  entanglement entropies}},\ }\href
  {https://doi.org/10.1103/PhysRevE.90.013308} {\bibfield  {journal} {\bibinfo
  {journal} {Phys. Rev. E}\ }\textbf {\bibinfo {volume} {90}},\ \bibinfo
  {pages} {013308} (\bibinfo {year} {2014}{\natexlab{a}})}\BibitemShut
  {NoStop}%
\bibitem [{\citenamefont {Kitaev}\ and\ \citenamefont
  {Preskill}(2006)}]{Kitaev2006}%
  \BibitemOpen
  \bibfield  {author} {\bibinfo {author} {\bibfnamefont {A.}~\bibnamefont
  {Kitaev}}\ and\ \bibinfo {author} {\bibfnamefont {J.}~\bibnamefont
  {Preskill}},\ }\bibfield  {title} {\bibinfo {title} {{Topological
  entanglement entropy}},\ }\href
  {https://doi.org/10.1103/PhysRevLett.96.110404} {\bibfield  {journal}
  {\bibinfo  {journal} {Phys. Rev. Lett.}\ }\textbf {\bibinfo {volume} {96}},\
  \bibinfo {pages} {110404} (\bibinfo {year} {2006})}\BibitemShut {NoStop}%
\bibitem [{\citenamefont {Islam}\ \emph {et~al.}(2015)\citenamefont {Islam},
  \citenamefont {Ma}, \citenamefont {Preiss}, \citenamefont {Tai},
  \citenamefont {Lukin}, \citenamefont {Rispoli},\ and\ \citenamefont
  {Greiner}}]{Islam2015}%
  \BibitemOpen
  \bibfield  {author} {\bibinfo {author} {\bibfnamefont {R.}~\bibnamefont
  {Islam}}, \bibinfo {author} {\bibfnamefont {R.}~\bibnamefont {Ma}}, \bibinfo
  {author} {\bibfnamefont {P.~M.}\ \bibnamefont {Preiss}}, \bibinfo {author}
  {\bibfnamefont {M.~E.}\ \bibnamefont {Tai}}, \bibinfo {author} {\bibfnamefont
  {A.}~\bibnamefont {Lukin}}, \bibinfo {author} {\bibfnamefont
  {M.}~\bibnamefont {Rispoli}},\ and\ \bibinfo {author} {\bibfnamefont
  {M.}~\bibnamefont {Greiner}},\ }\bibfield  {title} {\bibinfo {title}
  {{Measuring entanglement entropy in a quantum many-body system}},\ }\href
  {https://doi.org/10.1038/nature15750} {\bibfield  {journal} {\bibinfo
  {journal} {Nature}\ }\textbf {\bibinfo {volume} {528}},\ \bibinfo {pages}
  {77} (\bibinfo {year} {2015})}\BibitemShut {NoStop}%
\bibitem [{\citenamefont {D'Alessio}\ \emph {et~al.}(2016)\citenamefont
  {D'Alessio}, \citenamefont {Kafri}, \citenamefont {Polkovnikov},\ and\
  \citenamefont {Rigol}}]{DAlessio2016}%
  \BibitemOpen
  \bibfield  {author} {\bibinfo {author} {\bibfnamefont {L.}~\bibnamefont
  {D'Alessio}}, \bibinfo {author} {\bibfnamefont {Y.}~\bibnamefont {Kafri}},
  \bibinfo {author} {\bibfnamefont {A.}~\bibnamefont {Polkovnikov}},\ and\
  \bibinfo {author} {\bibfnamefont {M.}~\bibnamefont {Rigol}},\ }\bibfield
  {title} {\bibinfo {title} {{From quantum chaos and eigenstate thermalization
  to statistical mechanics and thermodynamics}},\ }\href
  {https://doi.org/10.1080/00018732.2016.1198134} {\bibfield  {journal}
  {\bibinfo  {journal} {Adv. Phys.}\ }\textbf {\bibinfo {volume} {65}},\
  \bibinfo {pages} {239} (\bibinfo {year} {2016})}\BibitemShut {NoStop}%
\bibitem [{\citenamefont {Nandkishore}\ and\ \citenamefont
  {Huse}(2015)}]{Nandkishore2015}%
  \BibitemOpen
  \bibfield  {author} {\bibinfo {author} {\bibfnamefont {R.}~\bibnamefont
  {Nandkishore}}\ and\ \bibinfo {author} {\bibfnamefont {D.~A.}\ \bibnamefont
  {Huse}},\ }\bibfield  {title} {\bibinfo {title} {{Many-body localization and
  thermalization in quantum statistical mechanics}},\ }\href
  {https://doi.org/10.1146/annurev-conmatphys-031214-014726} {\bibfield
  {journal} {\bibinfo  {journal} {Annu. Rev. Condens. Matter Phys.}\ }\textbf
  {\bibinfo {volume} {6}},\ \bibinfo {pages} {15} (\bibinfo {year}
  {2015})}\BibitemShut {NoStop}%
\bibitem [{\citenamefont {Alet}\ and\ \citenamefont
  {Laflorencie}(2018)}]{Alet2018}%
  \BibitemOpen
  \bibfield  {author} {\bibinfo {author} {\bibfnamefont {F.}~\bibnamefont
  {Alet}}\ and\ \bibinfo {author} {\bibfnamefont {N.}~\bibnamefont
  {Laflorencie}},\ }\bibfield  {title} {\bibinfo {title} {{Many-body
  localization: An introduction and selected topics}},\ }\href
  {https://doi.org/10.1016/j.crhy.2018.03.003} {\bibfield  {journal} {\bibinfo
  {journal} {Comptes Rendus Phys.}\ }\textbf {\bibinfo {volume} {19}},\
  \bibinfo {pages} {498} (\bibinfo {year} {2018})}\BibitemShut {NoStop}%
\bibitem [{\citenamefont {Brydges}\ \emph {et~al.}(2019)\citenamefont
  {Brydges}, \citenamefont {Elben}, \citenamefont {Jurcevic}, \citenamefont
  {Vermersch}, \citenamefont {Maier}, \citenamefont {Lanyon}, \citenamefont
  {Zoller}, \citenamefont {Blatt},\ and\ \citenamefont {Roos}}]{Brydges2020}%
  \BibitemOpen
  \bibfield  {author} {\bibinfo {author} {\bibfnamefont {T.}~\bibnamefont
  {Brydges}}, \bibinfo {author} {\bibfnamefont {A.}~\bibnamefont {Elben}},
  \bibinfo {author} {\bibfnamefont {P.}~\bibnamefont {Jurcevic}}, \bibinfo
  {author} {\bibfnamefont {B.}~\bibnamefont {Vermersch}}, \bibinfo {author}
  {\bibfnamefont {C.}~\bibnamefont {Maier}}, \bibinfo {author} {\bibfnamefont
  {B.~P.}\ \bibnamefont {Lanyon}}, \bibinfo {author} {\bibfnamefont
  {P.}~\bibnamefont {Zoller}}, \bibinfo {author} {\bibfnamefont
  {R.}~\bibnamefont {Blatt}},\ and\ \bibinfo {author} {\bibfnamefont {C.~F.}\
  \bibnamefont {Roos}},\ }\bibfield  {title} {\bibinfo {title} {{Probing
  R{\'{e}}nyi entanglement entropy via randomized measurements}},\ }\href
  {https://doi.org/10.1126/science.aau4963} {\bibfield  {journal} {\bibinfo
  {journal} {Science}\ }\textbf {\bibinfo {volume} {364}},\ \bibinfo {pages}
  {260} (\bibinfo {year} {2019})}\BibitemShut {NoStop}%
\bibitem [{\citenamefont {Herdman}\ \emph
  {et~al.}(2014{\natexlab{b}})\citenamefont {Herdman}, \citenamefont {Roy},
  \citenamefont {Melko},\ and\ \citenamefont {{Del Maestro}}}]{Herdman2014a}%
  \BibitemOpen
  \bibfield  {author} {\bibinfo {author} {\bibfnamefont {C.~M.}\ \bibnamefont
  {Herdman}}, \bibinfo {author} {\bibfnamefont {P.~N.}\ \bibnamefont {Roy}},
  \bibinfo {author} {\bibfnamefont {R.~G.}\ \bibnamefont {Melko}},\ and\
  \bibinfo {author} {\bibfnamefont {A.}~\bibnamefont {{Del Maestro}}},\
  }\bibfield  {title} {\bibinfo {title} {{Particle entanglement in continuum
  many-body systems via quantum Monte Carlo}},\ }\href
  {https://doi.org/10.1103/PhysRevB.89.140501} {\bibfield  {journal} {\bibinfo
  {journal} {Phys. Rev. B}\ }\textbf {\bibinfo {volume} {89}},\ \bibinfo
  {pages} {140501(R)} (\bibinfo {year} {2014}{\natexlab{b}})}\BibitemShut
  {NoStop}%
\bibitem [{\citenamefont {Zhao}\ \emph {et~al.}(2022)\citenamefont {Zhao},
  \citenamefont {Wang}, \citenamefont {Yan}, \citenamefont {Cheng},\ and\
  \citenamefont {Meng}}]{Zhao2022}%
  \BibitemOpen
  \bibfield  {author} {\bibinfo {author} {\bibfnamefont {J.}~\bibnamefont
  {Zhao}}, \bibinfo {author} {\bibfnamefont {Y.-c.}\ \bibnamefont {Wang}},
  \bibinfo {author} {\bibfnamefont {Z.}~\bibnamefont {Yan}}, \bibinfo {author}
  {\bibfnamefont {M.}~\bibnamefont {Cheng}},\ and\ \bibinfo {author}
  {\bibfnamefont {Z.~Y.}\ \bibnamefont {Meng}},\ }\bibfield  {title} {\bibinfo
  {title} {{Scaling of Entanglement Entropy at Deconfined Quantum
  Criticality}},\ }\href {https://doi.org/10.1103/PhysRevLett.128.010601}
  {\bibfield  {journal} {\bibinfo  {journal} {Phys. Rev. Lett.}\ }\textbf
  {\bibinfo {volume} {128}},\ \bibinfo {pages} {10601} (\bibinfo {year}
  {2022})}\BibitemShut {NoStop}%
\bibitem [{\citenamefont {Fan}\ \emph {et~al.}(2004)\citenamefont {Fan},
  \citenamefont {Korepin},\ and\ \citenamefont {Roychowdhury}}]{Fan2004}%
  \BibitemOpen
  \bibfield  {author} {\bibinfo {author} {\bibfnamefont {H.}~\bibnamefont
  {Fan}}, \bibinfo {author} {\bibfnamefont {V.}~\bibnamefont {Korepin}},\ and\
  \bibinfo {author} {\bibfnamefont {V.}~\bibnamefont {Roychowdhury}},\
  }\bibfield  {title} {\bibinfo {title} {{Entanglement in a Valence-Bond Solid
  State}},\ }\href {https://doi.org/10.1103/PhysRevLett.93.227203} {\bibfield
  {journal} {\bibinfo  {journal} {Phys. Rev. Lett.}\ }\textbf {\bibinfo
  {volume} {93}},\ \bibinfo {pages} {227203} (\bibinfo {year}
  {2004})}\BibitemShut {NoStop}%
\bibitem [{\citenamefont {Refael}\ and\ \citenamefont
  {Moore}(2004)}]{Refael2004}%
  \BibitemOpen
  \bibfield  {author} {\bibinfo {author} {\bibfnamefont {G.}~\bibnamefont
  {Refael}}\ and\ \bibinfo {author} {\bibfnamefont {J.~E.}\ \bibnamefont
  {Moore}},\ }\bibfield  {title} {\bibinfo {title} {{Entanglement Entropy of
  Random Quantum Critical Points in One Dimension}},\ }\href
  {https://doi.org/10.1103/PhysRevLett.93.260602} {\bibfield  {journal}
  {\bibinfo  {journal} {Phys. Rev. Lett.}\ }\textbf {\bibinfo {volume} {93}},\
  \bibinfo {pages} {260602} (\bibinfo {year} {2004})}\BibitemShut {NoStop}%
\bibitem [{\citenamefont {Franchini}(2007)}]{Franchini2007}%
  \BibitemOpen
  \bibfield  {author} {\bibinfo {author} {\bibfnamefont {F.}~\bibnamefont
  {Franchini}},\ }\bibfield  {title} {\bibinfo {title} {{Renyi entropy of the
  XY spin chain}},\ }\bibfield  {journal} {\bibinfo  {journal} {J. Phys. A}\
  }\textbf {\bibinfo {volume} {41}},\ \href
  {https://doi.org/10.1088/1751-8113/41/2/025302}
  {10.1088/1751-8113/41/2/025302} (\bibinfo {year} {2007})\BibitemShut
  {NoStop}%
\bibitem [{\citenamefont {Bertini}\ \emph {et~al.}(2019)\citenamefont
  {Bertini}, \citenamefont {Kos},\ and\ \citenamefont {Prosen}}]{Bertini2019}%
  \BibitemOpen
  \bibfield  {author} {\bibinfo {author} {\bibfnamefont {B.}~\bibnamefont
  {Bertini}}, \bibinfo {author} {\bibfnamefont {P.}~\bibnamefont {Kos}},\ and\
  \bibinfo {author} {\bibfnamefont {T.}~\bibnamefont {Prosen}},\ }\bibfield
  {title} {\bibinfo {title} {{Entanglement Spreading in a Minimal Model of
  Maximal Many-Body Quantum Chaos}},\ }\href
  {https://doi.org/10.1103/PhysRevX.9.021033} {\bibfield  {journal} {\bibinfo
  {journal} {Phys. Rev. X}\ }\textbf {\bibinfo {volume} {9}},\ \bibinfo {pages}
  {021033} (\bibinfo {year} {2019})}\BibitemShut {NoStop}%
\bibitem [{\citenamefont {Schuch}\ \emph {et~al.}(2008)\citenamefont {Schuch},
  \citenamefont {Wolf}, \citenamefont {Verstraete},\ and\ \citenamefont
  {Cirac}}]{Schuch2008}%
  \BibitemOpen
  \bibfield  {author} {\bibinfo {author} {\bibfnamefont {N.}~\bibnamefont
  {Schuch}}, \bibinfo {author} {\bibfnamefont {M.~M.}\ \bibnamefont {Wolf}},
  \bibinfo {author} {\bibfnamefont {F.}~\bibnamefont {Verstraete}},\ and\
  \bibinfo {author} {\bibfnamefont {J.~I.}\ \bibnamefont {Cirac}},\ }\bibfield
  {title} {\bibinfo {title} {{Entropy scaling and simulability by matrix
  product states}},\ }\href {https://doi.org/10.1103/PhysRevLett.100.030504}
  {\bibfield  {journal} {\bibinfo  {journal} {Phys. Rev. Lett.}\ }\textbf
  {\bibinfo {volume} {100}},\ \bibinfo {pages} {030504} (\bibinfo {year}
  {2008})}\BibitemShut {NoStop}%
\bibitem [{\citenamefont {Vidal}(2008)}]{Vidal2008}%
  \BibitemOpen
  \bibfield  {author} {\bibinfo {author} {\bibfnamefont {G.}~\bibnamefont
  {Vidal}},\ }\bibfield  {title} {\bibinfo {title} {{Class of quantum Many-Body
  states that can be efficiently simulated}},\ }\href
  {https://doi.org/10.1103/PhysRevLett.101.110501} {\bibfield  {journal}
  {\bibinfo  {journal} {Phys. Rev. Lett.}\ }\textbf {\bibinfo {volume} {101}},\
  \bibinfo {pages} {110501} (\bibinfo {year} {2008})}\BibitemShut {NoStop}%
\bibitem [{\citenamefont {Carleo}\ and\ \citenamefont
  {Troyer}(2017)}]{Carleo2017}%
  \BibitemOpen
  \bibfield  {author} {\bibinfo {author} {\bibfnamefont {G.}~\bibnamefont
  {Carleo}}\ and\ \bibinfo {author} {\bibfnamefont {M.}~\bibnamefont
  {Troyer}},\ }\bibfield  {title} {\bibinfo {title} {{Solving the quantum
  many-body problem with artificial neural networks}},\ }\href
  {https://doi.org/10.1126/science.aag2302} {\bibfield  {journal} {\bibinfo
  {journal} {Science}\ }\textbf {\bibinfo {volume} {355}},\ \bibinfo {pages}
  {602} (\bibinfo {year} {2017})}\BibitemShut {NoStop}%
\bibitem [{\citenamefont {Ceriotti}\ \emph {et~al.}(2013)\citenamefont
  {Ceriotti}, \citenamefont {Cuny}, \citenamefont {Parrinello},\ and\
  \citenamefont {Manolopoulos}}]{Ceriotti2013}%
  \BibitemOpen
  \bibfield  {author} {\bibinfo {author} {\bibfnamefont {M.}~\bibnamefont
  {Ceriotti}}, \bibinfo {author} {\bibfnamefont {J.}~\bibnamefont {Cuny}},
  \bibinfo {author} {\bibfnamefont {M.}~\bibnamefont {Parrinello}},\ and\
  \bibinfo {author} {\bibfnamefont {D.~E.}\ \bibnamefont {Manolopoulos}},\
  }\bibfield  {title} {\bibinfo {title} {{Nuclear quantum effects and hydrogen
  bond fluctuations in water}},\ }\href
  {https://doi.org/10.1073/pnas.1308560110} {\bibfield  {journal} {\bibinfo
  {journal} {Proc. Natl. Acad. Sci. U. S. A.}\ }\textbf {\bibinfo {volume}
  {110}},\ \bibinfo {pages} {15591} (\bibinfo {year} {2013})}\BibitemShut
  {NoStop}%
\bibitem [{\citenamefont {Drozdov}\ \emph {et~al.}(2019)\citenamefont
  {Drozdov}, \citenamefont {Kong}, \citenamefont {Minkov}, \citenamefont
  {Besedin}, \citenamefont {Kuzovnikov}, \citenamefont {Mozaffari},
  \citenamefont {Balicas}, \citenamefont {Balakirev}, \citenamefont {Graf},
  \citenamefont {Prakapenka}, \citenamefont {Greenberg}, \citenamefont
  {Knyazev}, \citenamefont {Tkacz},\ and\ \citenamefont
  {Eremets}}]{Drozdov2019}%
  \BibitemOpen
  \bibfield  {author} {\bibinfo {author} {\bibfnamefont {A.~P.}\ \bibnamefont
  {Drozdov}}, \bibinfo {author} {\bibfnamefont {P.~P.}\ \bibnamefont {Kong}},
  \bibinfo {author} {\bibfnamefont {V.~S.}\ \bibnamefont {Minkov}}, \bibinfo
  {author} {\bibfnamefont {S.~P.}\ \bibnamefont {Besedin}}, \bibinfo {author}
  {\bibfnamefont {M.~A.}\ \bibnamefont {Kuzovnikov}}, \bibinfo {author}
  {\bibfnamefont {S.}~\bibnamefont {Mozaffari}}, \bibinfo {author}
  {\bibfnamefont {L.}~\bibnamefont {Balicas}}, \bibinfo {author} {\bibfnamefont
  {F.~F.}\ \bibnamefont {Balakirev}}, \bibinfo {author} {\bibfnamefont {D.~E.}\
  \bibnamefont {Graf}}, \bibinfo {author} {\bibfnamefont {V.~B.}\ \bibnamefont
  {Prakapenka}}, \bibinfo {author} {\bibfnamefont {E.}~\bibnamefont
  {Greenberg}}, \bibinfo {author} {\bibfnamefont {D.~A.}\ \bibnamefont
  {Knyazev}}, \bibinfo {author} {\bibfnamefont {M.}~\bibnamefont {Tkacz}},\
  and\ \bibinfo {author} {\bibfnamefont {M.~I.}\ \bibnamefont {Eremets}},\
  }\bibfield  {title} {\bibinfo {title} {{Superconductivity at 250 K in
  lanthanum hydride under high pressures}},\ }\href
  {https://doi.org/10.1038/s41586-019-1201-8} {\bibfield  {journal} {\bibinfo
  {journal} {Nature}\ }\textbf {\bibinfo {volume} {569}},\ \bibinfo {pages}
  {528} (\bibinfo {year} {2019})}\BibitemShut {NoStop}%
\bibitem [{\citenamefont {Kong}\ \emph {et~al.}(2021)\citenamefont {Kong},
  \citenamefont {Minkov}, \citenamefont {Kuzovnikov}, \citenamefont {Drozdov},
  \citenamefont {Besedin}, \citenamefont {Mozaffari}, \citenamefont {Balicas},
  \citenamefont {Balakirev}, \citenamefont {Prakapenka}, \citenamefont
  {Chariton}, \citenamefont {Knyazev}, \citenamefont {Greenberg},\ and\
  \citenamefont {Eremets}}]{Kong2021}%
  \BibitemOpen
  \bibfield  {author} {\bibinfo {author} {\bibfnamefont {P.}~\bibnamefont
  {Kong}}, \bibinfo {author} {\bibfnamefont {V.~S.}\ \bibnamefont {Minkov}},
  \bibinfo {author} {\bibfnamefont {M.~A.}\ \bibnamefont {Kuzovnikov}},
  \bibinfo {author} {\bibfnamefont {A.~P.}\ \bibnamefont {Drozdov}}, \bibinfo
  {author} {\bibfnamefont {S.~P.}\ \bibnamefont {Besedin}}, \bibinfo {author}
  {\bibfnamefont {S.}~\bibnamefont {Mozaffari}}, \bibinfo {author}
  {\bibfnamefont {L.}~\bibnamefont {Balicas}}, \bibinfo {author} {\bibfnamefont
  {F.~F.}\ \bibnamefont {Balakirev}}, \bibinfo {author} {\bibfnamefont {V.~B.}\
  \bibnamefont {Prakapenka}}, \bibinfo {author} {\bibfnamefont
  {S.}~\bibnamefont {Chariton}}, \bibinfo {author} {\bibfnamefont {D.~A.}\
  \bibnamefont {Knyazev}}, \bibinfo {author} {\bibfnamefont {E.}~\bibnamefont
  {Greenberg}},\ and\ \bibinfo {author} {\bibfnamefont {M.~I.}\ \bibnamefont
  {Eremets}},\ }\bibfield  {title} {\bibinfo {title} {{Superconductivity up to
  243 K in the Yttrium-Hydrogen system under high pressure}},\ }\href
  {https://doi.org/10.1038/s41467-021-25372-2} {\bibfield  {journal} {\bibinfo
  {journal} {Nat. Commun.}\ }\textbf {\bibinfo {volume} {12}},\ \bibinfo
  {pages} {5075} (\bibinfo {year} {2021})}\BibitemShut {NoStop}%
\bibitem [{\citenamefont {Drozdov}\ \emph {et~al.}(2015)\citenamefont
  {Drozdov}, \citenamefont {Eremets}, \citenamefont {Troyan}, \citenamefont
  {Ksenofontov},\ and\ \citenamefont {Shylin}}]{Drozdov2015}%
  \BibitemOpen
  \bibfield  {author} {\bibinfo {author} {\bibfnamefont {A.~P.}\ \bibnamefont
  {Drozdov}}, \bibinfo {author} {\bibfnamefont {M.~I.}\ \bibnamefont
  {Eremets}}, \bibinfo {author} {\bibfnamefont {I.~A.}\ \bibnamefont {Troyan}},
  \bibinfo {author} {\bibfnamefont {V.}~\bibnamefont {Ksenofontov}},\ and\
  \bibinfo {author} {\bibfnamefont {S.~I.}\ \bibnamefont {Shylin}},\ }\bibfield
   {title} {\bibinfo {title} {{Conventional superconductivity at 203 kelvin at
  high pressures in the sulfur hydride system}},\ }\href
  {https://doi.org/10.1038/nature14964} {\bibfield  {journal} {\bibinfo
  {journal} {Nature}\ }\textbf {\bibinfo {volume} {525}},\ \bibinfo {pages}
  {73} (\bibinfo {year} {2015})}\BibitemShut {NoStop}%
\bibitem [{\citenamefont {Fillaux}(2005)}]{Fillaux2005}%
  \BibitemOpen
  \bibfield  {author} {\bibinfo {author} {\bibfnamefont {F.}~\bibnamefont
  {Fillaux}},\ }\bibfield  {title} {\bibinfo {title} {Quantum entanglement and
  nonlocal proton transfer dynamics in dimers of formic acid and analogues},\
  }\href {https://doi.org/https://doi.org/10.1016/j.cplett.2005.04.069}
  {\bibfield  {journal} {\bibinfo  {journal} {Chem. Phys. Lett.}\ }\textbf
  {\bibinfo {volume} {408}},\ \bibinfo {pages} {302} (\bibinfo {year}
  {2005})}\BibitemShut {NoStop}%
\bibitem [{\citenamefont {Miura}\ \emph {et~al.}(1998)\citenamefont {Miura},
  \citenamefont {Tuckerman},\ and\ \citenamefont {Klein}}]{Miura2010}%
  \BibitemOpen
  \bibfield  {author} {\bibinfo {author} {\bibfnamefont {S.}~\bibnamefont
  {Miura}}, \bibinfo {author} {\bibfnamefont {M.~E.}\ \bibnamefont
  {Tuckerman}},\ and\ \bibinfo {author} {\bibfnamefont {M.~L.}\ \bibnamefont
  {Klein}},\ }\bibfield  {title} {\bibinfo {title} {{An ab initio path integral
  molecular dynamics study of double proton transfer in the formic acid
  dimer}},\ }\href {https://doi.org/10.1063/1.477147} {\bibfield  {journal}
  {\bibinfo  {journal} {J. Chem. Phys.}\ }\textbf {\bibinfo {volume} {109}},\
  \bibinfo {pages} {5290} (\bibinfo {year} {1998})}\BibitemShut {NoStop}%
\bibitem [{\citenamefont {Ivanov}\ \emph {et~al.}(2015)\citenamefont {Ivanov},
  \citenamefont {Grant},\ and\ \citenamefont {Marx}}]{Marx2015}%
  \BibitemOpen
  \bibfield  {author} {\bibinfo {author} {\bibfnamefont {S.~D.}\ \bibnamefont
  {Ivanov}}, \bibinfo {author} {\bibfnamefont {I.~M.}\ \bibnamefont {Grant}},\
  and\ \bibinfo {author} {\bibfnamefont {D.}~\bibnamefont {Marx}},\ }\bibfield
  {title} {\bibinfo {title} {{Quantum free energy landscapes from ab initio
  path integral metadynamics: Double proton transfer in the formic acid dimer
  is concerted but not correlated}},\ }\bibfield  {journal} {\bibinfo
  {journal} {J. Chem. Phys.}\ }\textbf {\bibinfo {volume} {143}},\ \href
  {https://doi.org/10.1063/1.4931052} {10.1063/1.4931052} (\bibinfo {year}
  {2015})\BibitemShut {NoStop}%
\bibitem [{\citenamefont {Ceriotti}\ \emph {et~al.}(2016)\citenamefont
  {Ceriotti}, \citenamefont {Fang}, \citenamefont {Kusalik}, \citenamefont
  {McKenzie}, \citenamefont {Michaelides}, \citenamefont {Morales},\ and\
  \citenamefont {Markland}}]{Ceriotti2016}%
  \BibitemOpen
  \bibfield  {author} {\bibinfo {author} {\bibfnamefont {M.}~\bibnamefont
  {Ceriotti}}, \bibinfo {author} {\bibfnamefont {W.}~\bibnamefont {Fang}},
  \bibinfo {author} {\bibfnamefont {P.~G.}\ \bibnamefont {Kusalik}}, \bibinfo
  {author} {\bibfnamefont {R.~H.}\ \bibnamefont {McKenzie}}, \bibinfo {author}
  {\bibfnamefont {A.}~\bibnamefont {Michaelides}}, \bibinfo {author}
  {\bibfnamefont {M.~A.}\ \bibnamefont {Morales}},\ and\ \bibinfo {author}
  {\bibfnamefont {T.~E.}\ \bibnamefont {Markland}},\ }\bibfield  {title}
  {\bibinfo {title} {{Nuclear Quantum Effects in Water and Aqueous Systems:
  Experiment, Theory, and Current Challenges}},\ }\href
  {https://doi.org/10.1021/acs.chemrev.5b00674} {\bibfield  {journal} {\bibinfo
   {journal} {Chem. Rev.}\ }\textbf {\bibinfo {volume} {116}},\ \bibinfo
  {pages} {7529} (\bibinfo {year} {2016})}\BibitemShut {NoStop}%
\bibitem [{\citenamefont {Pusuluk}\ \emph {et~al.}(2018)\citenamefont
  {Pusuluk}, \citenamefont {Torun},\ and\ \citenamefont
  {Deliduman}}]{Pusuluk2018}%
  \BibitemOpen
  \bibfield  {author} {\bibinfo {author} {\bibfnamefont {O.}~\bibnamefont
  {Pusuluk}}, \bibinfo {author} {\bibfnamefont {G.}~\bibnamefont {Torun}},\
  and\ \bibinfo {author} {\bibfnamefont {C.}~\bibnamefont {Deliduman}},\
  }\bibfield  {title} {\bibinfo {title} {{Quantum entanglement shared in
  hydrogen bonds and its usage as a resource in molecular recognition}},\
  }\href {https://doi.org/10.1142/S0217984918503086} {\bibfield  {journal}
  {\bibinfo  {journal} {Mod. Phys. Lett. B}\ }\textbf {\bibinfo {volume}
  {32}},\ \bibinfo {pages} {1850308} (\bibinfo {year} {2018})}\BibitemShut
  {NoStop}%
\bibitem [{\citenamefont {Amico}\ \emph {et~al.}(2008)\citenamefont {Amico},
  \citenamefont {Fazio}, \citenamefont {Osterloh},\ and\ \citenamefont
  {Vedral}}]{Amico2008}%
  \BibitemOpen
  \bibfield  {author} {\bibinfo {author} {\bibfnamefont {L.}~\bibnamefont
  {Amico}}, \bibinfo {author} {\bibfnamefont {R.}~\bibnamefont {Fazio}},
  \bibinfo {author} {\bibfnamefont {A.}~\bibnamefont {Osterloh}},\ and\
  \bibinfo {author} {\bibfnamefont {V.}~\bibnamefont {Vedral}},\ }\bibfield
  {title} {\bibinfo {title} {{Entanglement in many-body systems}},\ }\href
  {https://doi.org/10.1103/RevModPhys.80.517} {\bibfield  {journal} {\bibinfo
  {journal} {Rev. Mod. Phys.}\ }\textbf {\bibinfo {volume} {80}},\ \bibinfo
  {pages} {517} (\bibinfo {year} {2008})}\BibitemShut {NoStop}%
\bibitem [{\citenamefont {Kinz-Thompson}\ and\ \citenamefont
  {Conwell}(2010)}]{Chen2009}%
  \BibitemOpen
  \bibfield  {author} {\bibinfo {author} {\bibfnamefont {C.}~\bibnamefont
  {Kinz-Thompson}}\ and\ \bibinfo {author} {\bibfnamefont {E.}~\bibnamefont
  {Conwell}},\ }\bibfield  {title} {\bibinfo {title} {{Proton transfer in
  adenine-thymine radical cation embedded in B-form DNA}},\ }\href
  {https://doi.org/10.1021/jz100214h} {\bibfield  {journal} {\bibinfo
  {journal} {J. Phys. Chem. Lett.}\ }\textbf {\bibinfo {volume} {1}},\ \bibinfo
  {pages} {1403} (\bibinfo {year} {2010})}\BibitemShut {NoStop}%
\bibitem [{\citenamefont {Ceperley}(1995)}]{Ceperley1995}%
  \BibitemOpen
  \bibfield  {author} {\bibinfo {author} {\bibfnamefont {D.~M.}\ \bibnamefont
  {Ceperley}},\ }\bibfield  {title} {\bibinfo {title} {{Path integrals in the
  theory of condensed helium}},\ }\href
  {https://doi.org/10.1103/RevModPhys.67.279} {\bibfield  {journal} {\bibinfo
  {journal} {Rev. Mod. Phys.}\ }\textbf {\bibinfo {volume} {67}},\ \bibinfo
  {pages} {279} (\bibinfo {year} {1995})}\BibitemShut {NoStop}%
\bibitem [{\citenamefont {Tuckerman}(2010)}]{Tuckerman}%
  \BibitemOpen
  \bibfield  {author} {\bibinfo {author} {\bibfnamefont {M.}~\bibnamefont
  {Tuckerman}},\ }\href@noop {} {\emph {\bibinfo {title} {{Statistical
  Mechanics: Theory and Molecular Simulation}}}}\ (\bibinfo  {publisher}
  {Oxford University Press Inc., New York},\ \bibinfo {year}
  {2010})\BibitemShut {NoStop}%
\bibitem [{\citenamefont {Chipot}\ and\ \citenamefont
  {Pohorille}(2007)}]{FreeEnergyBook}%
  \BibitemOpen
  \bibfield  {author} {\bibinfo {author} {\bibfnamefont {C.}~\bibnamefont
  {Chipot}}\ and\ \bibinfo {author} {\bibfnamefont {A.}~\bibnamefont
  {Pohorille}},\ }\href {https://doi.org/10.1007/978-3-540-38448-9} {\emph
  {\bibinfo {title} {{Free Energy Calculations}}}}\ (\bibinfo  {publisher}
  {Springer-Verlag Berlin Heidelberg},\ \bibinfo {year} {2007})\BibitemShut
  {NoStop}%
\bibitem [{\citenamefont {Bennett}(1976)}]{Bennett1976}%
  \BibitemOpen
  \bibfield  {author} {\bibinfo {author} {\bibfnamefont {C.~H.}\ \bibnamefont
  {Bennett}},\ }\bibfield  {title} {\bibinfo {title} {{Efficient Estimation of
  Free Energy Differences from Monte Carlo Data}},\ }\href
  {https://doi.org/10.1016/0021-9991(76)90078-4} {\bibfield  {journal}
  {\bibinfo  {journal} {J. Comput. Phys.}\ }\textbf {\bibinfo {volume} {268}},\
  \bibinfo {pages} {245} (\bibinfo {year} {1976})}\BibitemShut {NoStop}%
\bibitem [{\citenamefont {Broecker}\ and\ \citenamefont
  {Trebst}(2014)}]{Broecker2014}%
  \BibitemOpen
  \bibfield  {author} {\bibinfo {author} {\bibfnamefont {P.}~\bibnamefont
  {Broecker}}\ and\ \bibinfo {author} {\bibfnamefont {S.}~\bibnamefont
  {Trebst}},\ }\bibfield  {title} {\bibinfo {title} {{R{\'{e}}nyi entropies of
  interacting fermions from determinantal quantum Monte Carlo simulations}},\
  }\href {https://doi.org/10.1088/1742-5468/2014/08/p08015} {\bibfield
  {journal} {\bibinfo  {journal} {J. Stat. Mech. Theory Exp.}\ }\textbf
  {\bibinfo {volume} {2014}},\ \bibinfo {pages} {08015} (\bibinfo {year}
  {2014})}\BibitemShut {NoStop}%
\bibitem [{\citenamefont {Buividovich}\ and\ \citenamefont
  {Polikarpov}(2008)}]{Buividovich2008}%
  \BibitemOpen
  \bibfield  {author} {\bibinfo {author} {\bibfnamefont {P.~V.}\ \bibnamefont
  {Buividovich}}\ and\ \bibinfo {author} {\bibfnamefont {M.~I.}\ \bibnamefont
  {Polikarpov}},\ }\bibfield  {title} {\bibinfo {title} {{Numerical study of
  entanglement entropy in SU(2) lattice gauge theory}},\ }\href
  {https://doi.org/10.1016/j.nuclphysb.2008.04.024} {\bibfield  {journal}
  {\bibinfo  {journal} {Nucl. Phys. B}\ }\textbf {\bibinfo {volume} {802}},\
  \bibinfo {pages} {458} (\bibinfo {year} {2008})}\BibitemShut {NoStop}%
\bibitem [{SM(2021)}]{SM}%
  \BibitemOpen
  \href@noop {} {}\bibinfo {howpublished} {See Supplemental Material at
  [\emph{URL will be inserted by publisher}] for additional information about
  the computational details for the quantum Harmonic oscillator, the
  one-dimensional Ising model, and the \emph{ab initio} calculations of the
  formic acid dimer. \cite{Krzakala2008, Deng2002, Mbeng2020, FreeEnergyBook,
  Sun2018, Sun2020}} (\bibinfo {year} {2021})\BibitemShut {NoStop}%
\bibitem [{\citenamefont {Mbeng}\ \emph {et~al.}(2020)\citenamefont {Mbeng},
  \citenamefont {Russomanno},\ and\ \citenamefont {Santoro}}]{Mbeng2020}%
  \BibitemOpen
  \bibfield  {author} {\bibinfo {author} {\bibfnamefont {G.~B.}\ \bibnamefont
  {Mbeng}}, \bibinfo {author} {\bibfnamefont {A.}~\bibnamefont {Russomanno}},\
  and\ \bibinfo {author} {\bibfnamefont {G.~E.}\ \bibnamefont {Santoro}},\
  }\href@noop {} {\bibinfo {title} {{The quantum Ising chain for beginners}}}
  (\bibinfo {year} {2020}),\ \Eprint {https://arxiv.org/abs/2009.09208v1}
  {arXiv:2009.09208v1} \BibitemShut {NoStop}%
\bibitem [{Srd(2021)}]{Srdinsek2021github}%
  \BibitemOpen
  \href {https://github.com/srdinsek/Renyi-Integration} {}\bibinfo
  {howpublished} {The code used in this work may be accessed at
  \url{https://github.com/srdinsek/Renyi-Integration}.} (\bibinfo {year}
  {2021})\BibitemShut {NoStop}%
\bibitem [{\citenamefont {Tachikawa}(2020)}]{Tachikawa2020}%
  \BibitemOpen
  \bibfield  {author} {\bibinfo {author} {\bibfnamefont {H.}~\bibnamefont
  {Tachikawa}},\ }\bibfield  {title} {\bibinfo {title} {{Proton Transfer vs
  Complex Formation Channels in Ionized Formic Acid Dimer: A Direct Ab Initio
  Molecular Dynamics Study}},\ }\href
  {https://doi.org/10.1021/acs.jpca.0c01729} {\bibfield  {journal} {\bibinfo
  {journal} {J. Phys. Chem. A}\ }\textbf {\bibinfo {volume} {124}},\ \bibinfo
  {pages} {3048} (\bibinfo {year} {2020})}\BibitemShut {NoStop}%
\bibitem [{\citenamefont {Wilde}(2011)}]{Wilde2011}%
  \BibitemOpen
  \bibfield  {author} {\bibinfo {author} {\bibfnamefont {M.~M.}\ \bibnamefont
  {Wilde}},\ }\href {https://doi.org/10.1017/9781316809976.001} {\emph
  {\bibinfo {title} {{From Classical to Quantum Shannon Theory}}}}\ (\bibinfo
  {publisher} {Cambridge University Press},\ \bibinfo {year}
  {2011})\BibitemShut {NoStop}%
\bibitem [{\citenamefont {Krzakala}\ \emph {et~al.}(2008)\citenamefont
  {Krzakala}, \citenamefont {Rosso}, \citenamefont {Semerjian},\ and\
  \citenamefont {Zamponi}}]{Krzakala2008}%
  \BibitemOpen
  \bibfield  {author} {\bibinfo {author} {\bibfnamefont {F.}~\bibnamefont
  {Krzakala}}, \bibinfo {author} {\bibfnamefont {A.}~\bibnamefont {Rosso}},
  \bibinfo {author} {\bibfnamefont {G.}~\bibnamefont {Semerjian}},\ and\
  \bibinfo {author} {\bibfnamefont {F.}~\bibnamefont {Zamponi}},\ }\bibfield
  {title} {\bibinfo {title} {{Path-integral representation for quantum spin
  models: Application to the quantum cavity method and Monte Carlo
  simulations}},\ }\href {https://doi.org/10.1103/PhysRevB.78.134428}
  {\bibfield  {journal} {\bibinfo  {journal} {Phys. Rev. B}\ }\textbf {\bibinfo
  {volume} {78}},\ \bibinfo {pages} {134428} (\bibinfo {year}
  {2008})}\BibitemShut {NoStop}%
\bibitem [{\citenamefont {Bl{\"{o}}te}\ and\ \citenamefont
  {Deng}(2002)}]{Deng2002}%
  \BibitemOpen
  \bibfield  {author} {\bibinfo {author} {\bibfnamefont {H.~W.~J.}\
  \bibnamefont {Bl{\"{o}}te}}\ and\ \bibinfo {author} {\bibfnamefont
  {Y.}~\bibnamefont {Deng}},\ }\bibfield  {title} {\bibinfo {title} {{Cluster
  Monte Carlo simulation of the transverse Ising model}},\ }\href
  {https://doi.org/10.1103/PhysRevE.66.066110} {\bibfield  {journal} {\bibinfo
  {journal} {Phys. Rev. E}\ }\textbf {\bibinfo {volume} {66}},\ \bibinfo
  {pages} {066110} (\bibinfo {year} {2002})}\BibitemShut {NoStop}%
\bibitem [{\citenamefont {Sun}\ \emph {et~al.}(2018)\citenamefont {Sun},
  \citenamefont {Berkelbach}, \citenamefont {Blunt}, \citenamefont {Booth},
  \citenamefont {Guo}, \citenamefont {Li}, \citenamefont {Liu}, \citenamefont
  {McClain}, \citenamefont {Sayfutyarova}, \citenamefont {Sharma},
  \citenamefont {Wouters},\ and\ \citenamefont {Chan}}]{Sun2018}%
  \BibitemOpen
  \bibfield  {author} {\bibinfo {author} {\bibfnamefont {Q.}~\bibnamefont
  {Sun}}, \bibinfo {author} {\bibfnamefont {T.~C.}\ \bibnamefont {Berkelbach}},
  \bibinfo {author} {\bibfnamefont {N.~S.}\ \bibnamefont {Blunt}}, \bibinfo
  {author} {\bibfnamefont {G.~H.}\ \bibnamefont {Booth}}, \bibinfo {author}
  {\bibfnamefont {S.}~\bibnamefont {Guo}}, \bibinfo {author} {\bibfnamefont
  {Z.}~\bibnamefont {Li}}, \bibinfo {author} {\bibfnamefont {J.}~\bibnamefont
  {Liu}}, \bibinfo {author} {\bibfnamefont {J.~D.}\ \bibnamefont {McClain}},
  \bibinfo {author} {\bibfnamefont {E.~R.}\ \bibnamefont {Sayfutyarova}},
  \bibinfo {author} {\bibfnamefont {S.}~\bibnamefont {Sharma}}, \bibinfo
  {author} {\bibfnamefont {S.}~\bibnamefont {Wouters}},\ and\ \bibinfo {author}
  {\bibfnamefont {G.~K.~L.}\ \bibnamefont {Chan}},\ }\bibfield  {title}
  {\bibinfo {title} {{PySCF: the Python-based simulations of chemistry
  framework}},\ }\href {https://doi.org/10.1002/wcms.1340} {\bibfield
  {journal} {\bibinfo  {journal} {Wiley Interdiscip. Rev. Comput. Mol. Sci.}\
  }\textbf {\bibinfo {volume} {8}},\ \bibinfo {pages} {1} (\bibinfo {year}
  {2018})}\BibitemShut {NoStop}%
\bibitem [{\citenamefont {Sun}\ \emph {et~al.}(2020)\citenamefont {Sun},
  \citenamefont {Zhang}, \citenamefont {Banerjee}, \citenamefont {Bao},
  \citenamefont {Barbry}, \citenamefont {Blunt}, \citenamefont {Bogdanov},
  \citenamefont {Booth}, \citenamefont {Chen}, \citenamefont {Cui},
  \citenamefont {Eriksen}, \citenamefont {Gao}, \citenamefont {Guo},
  \citenamefont {Hermann}, \citenamefont {Hermes}, \citenamefont {Koh},
  \citenamefont {Koval}, \citenamefont {Lehtola}, \citenamefont {Li},
  \citenamefont {Liu}, \citenamefont {Mardirossian}, \citenamefont {McClain},
  \citenamefont {Motta}, \citenamefont {Mussard}, \citenamefont {Pham},
  \citenamefont {Pulkin}, \citenamefont {Purwanto}, \citenamefont {Robinson},
  \citenamefont {Ronca}, \citenamefont {Sayfutyarova}, \citenamefont
  {Scheurer}, \citenamefont {Schurkus}, \citenamefont {Smith}, \citenamefont
  {Sun}, \citenamefont {Sun}, \citenamefont {Upadhyay}, \citenamefont {Wagner},
  \citenamefont {Wang}, \citenamefont {White}, \citenamefont {Whitfield},
  \citenamefont {Williamson}, \citenamefont {Wouters}, \citenamefont {Yang},
  \citenamefont {Yu}, \citenamefont {Zhu}, \citenamefont {Berkelbach},
  \citenamefont {Sharma}, \citenamefont {Sokolov},\ and\ \citenamefont
  {Chan}}]{Sun2020}%
  \BibitemOpen
  \bibfield  {author} {\bibinfo {author} {\bibfnamefont {Q.}~\bibnamefont
  {Sun}}, \bibinfo {author} {\bibfnamefont {X.}~\bibnamefont {Zhang}}, \bibinfo
  {author} {\bibfnamefont {S.}~\bibnamefont {Banerjee}}, \bibinfo {author}
  {\bibfnamefont {P.}~\bibnamefont {Bao}}, \bibinfo {author} {\bibfnamefont
  {M.}~\bibnamefont {Barbry}}, \bibinfo {author} {\bibfnamefont {N.~S.}\
  \bibnamefont {Blunt}}, \bibinfo {author} {\bibfnamefont {N.~A.}\ \bibnamefont
  {Bogdanov}}, \bibinfo {author} {\bibfnamefont {G.~H.}\ \bibnamefont {Booth}},
  \bibinfo {author} {\bibfnamefont {J.}~\bibnamefont {Chen}}, \bibinfo {author}
  {\bibfnamefont {Z.~H.}\ \bibnamefont {Cui}}, \bibinfo {author} {\bibfnamefont
  {J.~J.}\ \bibnamefont {Eriksen}}, \bibinfo {author} {\bibfnamefont
  {Y.}~\bibnamefont {Gao}}, \bibinfo {author} {\bibfnamefont {S.}~\bibnamefont
  {Guo}}, \bibinfo {author} {\bibfnamefont {J.}~\bibnamefont {Hermann}},
  \bibinfo {author} {\bibfnamefont {M.~R.}\ \bibnamefont {Hermes}}, \bibinfo
  {author} {\bibfnamefont {K.}~\bibnamefont {Koh}}, \bibinfo {author}
  {\bibfnamefont {P.}~\bibnamefont {Koval}}, \bibinfo {author} {\bibfnamefont
  {S.}~\bibnamefont {Lehtola}}, \bibinfo {author} {\bibfnamefont
  {Z.}~\bibnamefont {Li}}, \bibinfo {author} {\bibfnamefont {J.}~\bibnamefont
  {Liu}}, \bibinfo {author} {\bibfnamefont {N.}~\bibnamefont {Mardirossian}},
  \bibinfo {author} {\bibfnamefont {J.~D.}\ \bibnamefont {McClain}}, \bibinfo
  {author} {\bibfnamefont {M.}~\bibnamefont {Motta}}, \bibinfo {author}
  {\bibfnamefont {B.}~\bibnamefont {Mussard}}, \bibinfo {author} {\bibfnamefont
  {H.~Q.}\ \bibnamefont {Pham}}, \bibinfo {author} {\bibfnamefont
  {A.}~\bibnamefont {Pulkin}}, \bibinfo {author} {\bibfnamefont
  {W.}~\bibnamefont {Purwanto}}, \bibinfo {author} {\bibfnamefont {P.~J.}\
  \bibnamefont {Robinson}}, \bibinfo {author} {\bibfnamefont {E.}~\bibnamefont
  {Ronca}}, \bibinfo {author} {\bibfnamefont {E.~R.}\ \bibnamefont
  {Sayfutyarova}}, \bibinfo {author} {\bibfnamefont {M.}~\bibnamefont
  {Scheurer}}, \bibinfo {author} {\bibfnamefont {H.~F.}\ \bibnamefont
  {Schurkus}}, \bibinfo {author} {\bibfnamefont {J.~E.}\ \bibnamefont {Smith}},
  \bibinfo {author} {\bibfnamefont {C.}~\bibnamefont {Sun}}, \bibinfo {author}
  {\bibfnamefont {S.~N.}\ \bibnamefont {Sun}}, \bibinfo {author} {\bibfnamefont
  {S.}~\bibnamefont {Upadhyay}}, \bibinfo {author} {\bibfnamefont {L.~K.}\
  \bibnamefont {Wagner}}, \bibinfo {author} {\bibfnamefont {X.}~\bibnamefont
  {Wang}}, \bibinfo {author} {\bibfnamefont {A.}~\bibnamefont {White}},
  \bibinfo {author} {\bibfnamefont {J.~D.}\ \bibnamefont {Whitfield}}, \bibinfo
  {author} {\bibfnamefont {M.~J.}\ \bibnamefont {Williamson}}, \bibinfo
  {author} {\bibfnamefont {S.}~\bibnamefont {Wouters}}, \bibinfo {author}
  {\bibfnamefont {J.}~\bibnamefont {Yang}}, \bibinfo {author} {\bibfnamefont
  {J.~M.}\ \bibnamefont {Yu}}, \bibinfo {author} {\bibfnamefont
  {T.}~\bibnamefont {Zhu}}, \bibinfo {author} {\bibfnamefont {T.~C.}\
  \bibnamefont {Berkelbach}}, \bibinfo {author} {\bibfnamefont
  {S.}~\bibnamefont {Sharma}}, \bibinfo {author} {\bibfnamefont {A.~Y.}\
  \bibnamefont {Sokolov}},\ and\ \bibinfo {author} {\bibfnamefont {G.~K.~L.}\
  \bibnamefont {Chan}},\ }\bibfield  {title} {\bibinfo {title} {{Recent
  developments in the PySCF program package}},\ }\href
  {https://doi.org/10.1063/5.0006074} {\bibfield  {journal} {\bibinfo
  {journal} {J. Chem. Phys.}\ }\textbf {\bibinfo {volume} {153}},\ \bibinfo
  {pages} {1} (\bibinfo {year} {2020})}\BibitemShut {NoStop}%
\end{thebibliography}%

\setcounter{equation}{0}
\setcounter{figure}{0}
\setcounter{table}{0}
\makeatletter
\renewcommand{\theequation}{S\arabic{equation}}
\renewcommand{\thefigure}{S\arabic{figure}}
\renewcommand{\bibnumfmt}[1]{[S#1]}

\onecolumngrid

\newpage
\section*{Quantum R\'enyi entropy by optimal thermodynamic integration paths\\Supplemental Material}

The supplemental material is organized as follows.
In Sec.~\ref{oscillator} we introduce the computational details for the quantum Harmonic oscillator, in Sec.~\ref{ising} we explain how we simulated the one-dimensional Ising model, while in Sec.~\ref{formic} we detail some numerical aspects of the \emph{ab initio} calculations of the formic acid dimer.
\vskip\baselineskip
\twocolumngrid

\section{Harmonic oscillator}
\label{oscillator}

The one-dimensional (1D) quantum harmonic oscillator was used as
the minimal model that retains the pathological behavior of the integrand at the edges of
the thermodynamic integration in Eq.~3 of the main text.
The expression in Eq.~3, proportional to the R\'enyi entropy, can be recast in the following form:
\begin{equation}
  \log\Big(\frac{\mathcal{Z}_{A}}{ \mathcal{Z}_{\emptyset}}\Big)=  -\beta \int_0^1 \boldsymbol{p'}\cdot\boldsymbol{K}  d\lambda,
    \label{eq:renyi_recast}
\end{equation}
where the components of $\boldsymbol{K} \equiv (\langle K_{\emptyset}\rangle_{\lambda},\langle K_A\rangle_{\lambda})$ are the energies of the springs connecting the beads at the boundaries of $\emptyset$ and $A$ ensembles, respectively, the quantum averages in $\boldsymbol{K}$ are taken over the the distribution generated by $H(g,h)$, and the integral is performed over the path $\boldsymbol{p}:[0, 1]\to\mathbb{R}^2$, a differentiable curve defined as $(g,h)(\lambda)=((1 - \lambda)^l,\lambda^l)$. $\boldsymbol{p'} \equiv (\partial g/\partial \lambda, \partial h/\partial \lambda)$ is the path direction in the $(g,h)$ plane.

\begin{figure}[b]
\includegraphics[width=1\linewidth]{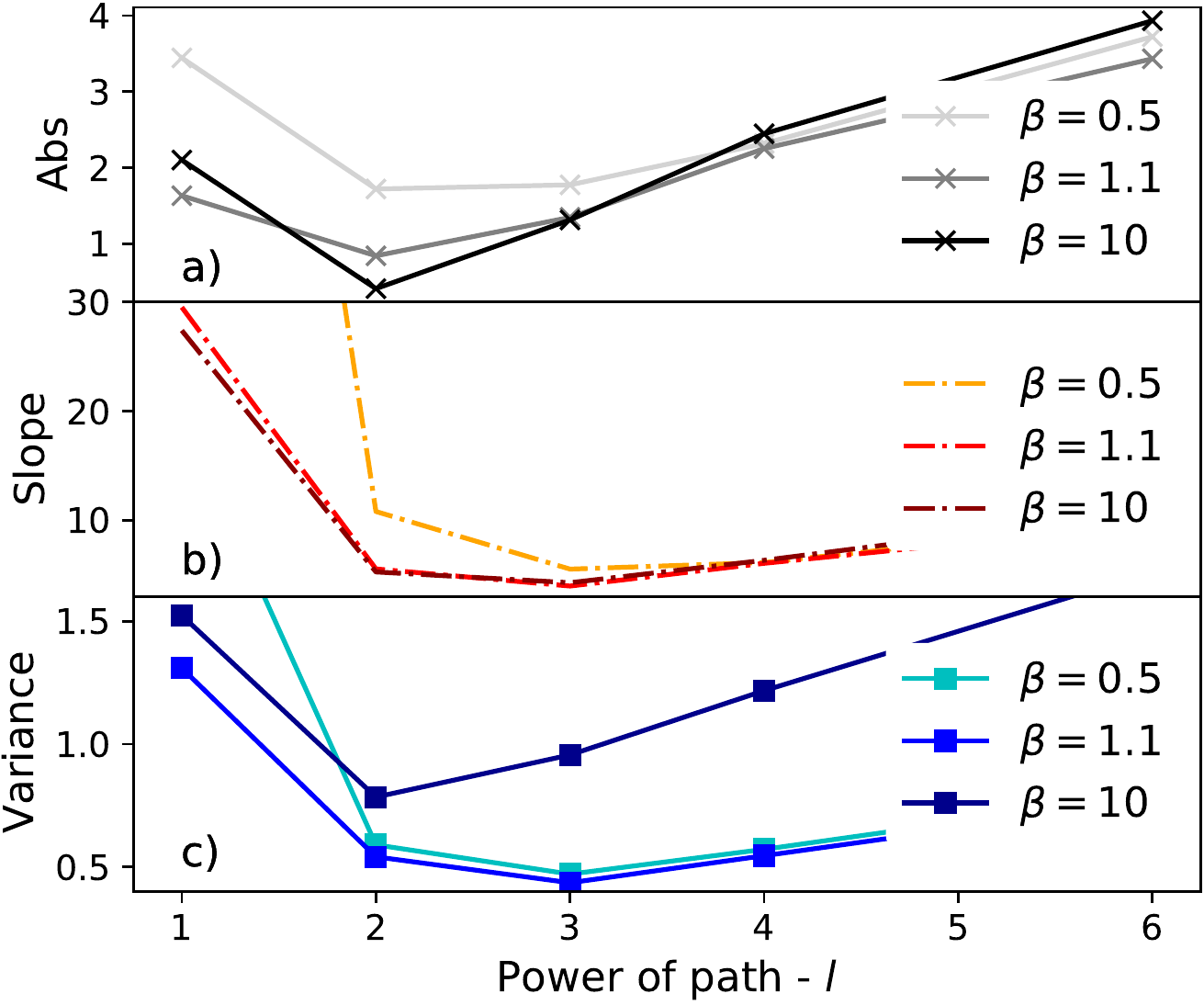}
\caption{\label{Appfig:Appendix_OptimalPath}
Cost functionals estimating a) excess area under the integrand ($F_\textrm{Abs}$ in Eq.~\ref{eq:functional1}), b) deviations from a linear integrand ($F_\textrm{Slope}$ in Eq.~\ref{eq:functional2}) and c) variance of the integral ($F_\textrm{Variance}$ in Eq.~\ref{eq:functional3}) for different regularizing paths defined by the parameter $l$. Three values of the temperature are shown. The lighter the color, the higher the temperature. 
}
\end{figure}

To find the optimal shape of the path that best regularizes the integrand in this system, we evaluated three cost functionals:
\begin{eqnarray}
F_\textrm{Abs}[\boldsymbol{p}] = \frac{1}{\int |\boldsymbol{p'}|d\lambda}\int_0^1 |\boldsymbol{p'}\cdot\boldsymbol{K}|d\lambda,
\label{eq:functional1}
\\
F_\textrm{Slope}[\boldsymbol{p}] = \frac{1}{\int |\boldsymbol{p'}|d\lambda}\int_0^1 |\partial_\lambda(\boldsymbol{p'}\cdot\boldsymbol{K})|d\lambda,
\label{eq:functional2}
\\
F_\textrm{Variance}[\boldsymbol{p}] = \int_0^1 \boldsymbol{p'}^2\cdot\text{var}[\boldsymbol{K}]d\lambda.
\label{eq:functional3}
\end{eqnarray}
In the above expressions, $\int |\boldsymbol{p'}|d\lambda$ is the length of the full path, $\boldsymbol{p'}^2 \equiv ((\partial g/\partial \lambda)^2, (\partial h/\partial \lambda)^2)$, and $\text{var}[\boldsymbol{K}]\equiv (\text{var}[\langle K_{\emptyset}\rangle_{\lambda}],\text{var}[\langle K_A\rangle_{\lambda}])$ is the variance of both $\boldsymbol{K}$ components.

Since the integration in Eq.~\ref{eq:renyi_recast} 
depends only on the initial $(g,h)(0)=(1,0)$ and final $(g,h)(1)=(0,1)$ points of the path $\boldsymbol{p}$,
the behavior
of the integrand can be tested with the 
functional $F_\textrm{Abs}$ in Eq.~\ref{eq:functional1} 
and Fig.~\ref{Appfig:Appendix_OptimalPath}a. 
This functional will disclose paths that 
lead to cancellation between large contributions of opposite sign.
Indeed, it will reach its minimal (and optimal) value when its positive and negative contributions are minimized in absolute value. 
With the aim at performing an efficient integration,
not only the value but
also the shape of the integrand is relevant. Indeed, an oscillating integrand requires more integration steps. The functional $F_\textrm{Slope}$ in Eq.~\ref{eq:functional2} 
and Fig.~\ref{Appfig:Appendix_OptimalPath}b 
sums up
the absolute value of the integrand derivative along the path and reaches the minimum when the integrand is linear.
Finally, the 
functional $F_\textrm{Variance}$ in Eq.~\ref{eq:functional3} 
and Fig.~\ref{Appfig:Appendix_OptimalPath}c 
is
the variance of the integral, assuming no covariance between $K_{\emptyset}$ and $K_A$, for simplicity. 
It originates from the fluctuations in the stochastic evaluation of $\boldsymbol{K}$, which accumulate along the path. It is an estimate of the stochastic rather than deterministic uncertainty in the computation of the integral.

The smallest variance at 
high temperature ($\beta=0.5$)
is
reached when $l=3$ (Fig.~\ref{Appfig:Appendix_OptimalPath}c), but all the paths with $l>1$ 
are
substantially better than the linear one (with $l=1$). Very similar observations can be made for the slope functional (Fig.~\ref{Appfig:Appendix_OptimalPath}b).
This outcome is expected, given the behavior of the integrands
shown
in Figs.~2c-2e of the Letter. Indeed, at $l=3$, the integrand 
is
almost linear at low temperatures. 
While $F_\textrm{Slope}$ and $F_\textrm{Variance}$ agree in 
yielding
$l=3$ as best path, $F_\textrm{Abs}$ favors $l=2$ (see Fig.~\ref{Appfig:Appendix_OptimalPath}a),
because negative contributions to the integral almost vanish in that case. In $F_\textrm{Abs}$, the value of $l=3$ is, however, not much worse compared to $l=2$, particularly at high temperature. Thus, $l=3$ is the best compromise between the different cost functionals presented here and, therefore, it
is
the optimal power law to be used in the regularizing path, not only for the harmonic oscillator but also for more challenging realistic systems.

\section{Ising model}
\label{ising}

\subsection{Quantum-to-classical isomorphism}
The 1D quantum Ising model, studied in the Letter, was analysed using Path Integral Monte Carlo simulations with action expanded up to the second order in the Suzuki–Trotter breakup.
At this order of the breakup, 
the model can be formally interpreted as a two-dimensional (2D) classical anisotropic Ising model\cite{Krzakala2008} with 
Hamiltonian of the form
\begin{eqnarray}
H = \sum_{i}^N\sum_{j}^{P}\sigma_{i,j}^z\sigma_{i+1,j}^z -\frac{\log(\tanh(\zeta r))}{\zeta}\sigma_{i,j}^z\sigma_{i,j+1}^z,\qquad
\label{eq:IsingHamiltonian}
\end{eqnarray}
where $\zeta = \beta/ P$ is the inverse temperature of the classical analogue, $P$ is the number of beads and $N$ is the number of spins in the 1D quantum model. Only when $P\to\infty$ the equivalence is formally exact. For numerical investigations, we restrict ourselves to finite $P$ with error proportional to $\mathcal{O}(\beta^3/P^3)$. The first index 
refers to
the spin position in the quantum chain and the second one to the position in the imaginary-time axis. Each $\sigma_{i, j}$ is allowed to take values 
$\in \{-1, 1\}$. In the 
original 1D spin
model we used periodic boundary conditions,
which
translate into $\sigma^z_{N + 1, j} = \sigma^z_{1, j}$ in the classical model. $r$ is the strength of the magnetic field in the $x$ direction, and causes the interaction term to diverge when $r\to 0$.

\subsection{Sampling schemes}
We evaluated the R\'enyi entanglement entropy
by measuring the free energy cost of modifying the topology on a $N\times \alpha P$ grid,
by changing the boundary conditions in the imaginary-time direction. In the $\mathcal{Z}_{\emptyset}$ ensemble, 
they lead to imaginary-time indices defined as
\begin{eqnarray}
\sigma^z_{i, (\tau-1)P+f(j)} \;\; \text{for $\tau \in \{1,\ldots,\alpha\}$},
\label{eq:IsingBoundaryConditionsEmpty}
\end{eqnarray}
with $f:\mathbb{N} \rightarrow \mathbb{N}$ a function such that $f(i)=i$ for $i \in \{1, \ldots, P\}$ and $f(P+1)=1$,
which is a consequence of the trace. In the joint $\mathcal{Z}_{A}$ ensemble, where a subset $A$ of particles is chosen, the boundary conditions stay the same for the complement $i\in\bar{A}$, but change for $i_A\in A$ such that 
\begin{eqnarray}
\sigma^z_{i_A, \alpha P + 1}=\sigma^z_{i_A, 1}.
\label{eq:IsingBoundaryConditionsJoin}
\end{eqnarray}
This is a consequence of first raising the reduced density matrix to the power $\alpha$ and only 
then taking the trace over the subsystem. When the thermodynamic integration is 
performed,
both boundary conditions are 
used, 
but the interaction is weighted with parameters $g$ and $h$, defining the path (see Eq. 5 of the main text). 

The most efficient sampling scheme for this system is not the standard Metropolis-Hastings algorithm, but the cluster algorithm adapted to the anisotropic 2D classical Ising model\cite{Deng2002}, where 
all proposed moves are accepted.
Instead of flipping one spin or the whole row, a cluster of spins is created and each spin added with a probability that fulfills the detailed balance condition. After the cluster is constructed, all 
its spins
are flipped. The algorithm has to be slightly modified for the purpose of thermodynamic integration, where the spins lying on the boundary have three neighbours in the imaginary-time direction. The cluster can grow in both directions there, while the probability of 
growing in each direction
is controlled by the corresponding weights.

Evaluation of full entropy with the "swap" operator is inefficient when entropy is large, but this can be partially cured, by splitting the calculation in parts
\begin{eqnarray}
\frac{Z_{L}}{Z_{0}} = \frac{Z_{L}}{Z_{L-1}}\frac{Z_{L-1}}{Z_{L-2}}...\frac{Z_{2}}{Z_{1}}\frac{Z_{1}}{Z_{0}},
\label{eq:RatioTrick}
\end{eqnarray}
for $L$ equal to the subsystem size.
Every ratio can be 
computed
in a separate simulation and only 
at
the end the product is performed. 

In order to make a fair comparison between the 
thermodynamic integration with the path regulatization and the direct evaluation of the partition function ratios in Eq.~\ref{eq:RatioTrick},
the "swap" operator was split in 
a
number of steps 
equal to the one
in the thermodynamic integration, and the same sampling scheme was used,
with the same number of cluster algorithm steps.
We ran
16 integration steps and 8 intermediate points with "swap" operator, based on Metropolis–Hastings transition probabilities $\langle\text{min}(1, \exp(-\beta\Delta H))\rangle_{\mathcal{Z}_{\emptyset}} / \langle\text{min}(1, \exp(\beta \Delta H))\rangle_{\mathcal{Z}_{A}}$.
The results of this comparison are shown in Fig.~3 of the Letter.

Although it is not really 
necessary for the simulation of this system,
a multi-walker algorithm was used, where 
each walker follows its own Markov chain, and contributes to the total average.
The final error is evaluated as the standard deviation of the averages 
given
by each
walker. 

\begin{figure}[t]
\includegraphics[width=1\linewidth]{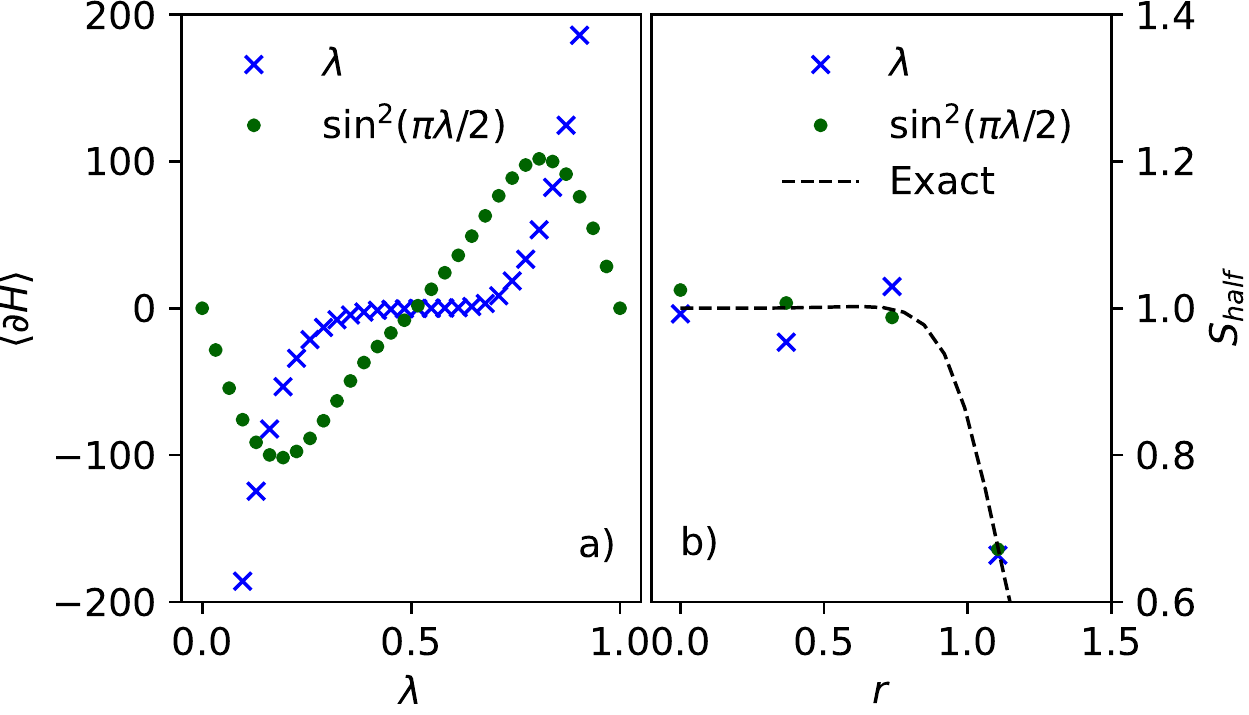}
\caption{\label{Appfig:Appendix_IsingModel}
a) Integrand $\langle \partial_\lambda H \rangle $ as a function of $\lambda$ in the Ising model at $\beta=5$ for the path regularization in Eq.~5 with 
$l=6$ (blue crosses) and for the same path but with rescaled $\lambda$, according to Eq.~\ref{eq:lambda_rescaling} (green points). 
b) Second order R\'enyi entanglement entropy computed for half of the system ($S_\textrm{half}$) as a function of $r$, for the two path regularizations, compared to the analytical result\cite{Mbeng2020}.
}
\end{figure}

\subsection{Different integration paths in 1D quantum Ising model}

The Ising model substantially differs from the \textit{ab initio} Hamiltonian already by the fact that spins are allowed to take only two values. This bounds the interaction in the imaginary-time direction and the divergence like the one observed in the harmonic oscillator does not appear. 
However,
the interaction strength still diverges with $r\to 0$, and 
a move 
connecting
two split ensembles is too costly
to be accepted. The regularizing path introduced in the Letter is therefore 
relevant again, because it 
reduces significantly
the variance and allows 
one
to produce 
accurate results up to very large system sizes.

Although 
the path regularization
with \miha{$l=3$}
leads to good results with 
low variance in the 1D Ising model (Fig.~\ref{Appfig:Appendix_IsingModel}b), one is still left with the freedom 
of adjusting
the density of points along the path, based on some \emph{quadrature rule}. Thus, more points can be added close to the difficult endpoints, resulting in a slower pace along the path and, hence, reducing the weight of integration steps with high-density points.
This reshapes the 
integrand
and can further reduce the integration error \cite{FreeEnergyBook}. The drawback of this approach is that a denser integration grid could lead to an increased computational cost, if there is no particular gain
in making
the integrand smoother.

\begin{figure}[b]
\includegraphics[width=1\linewidth]{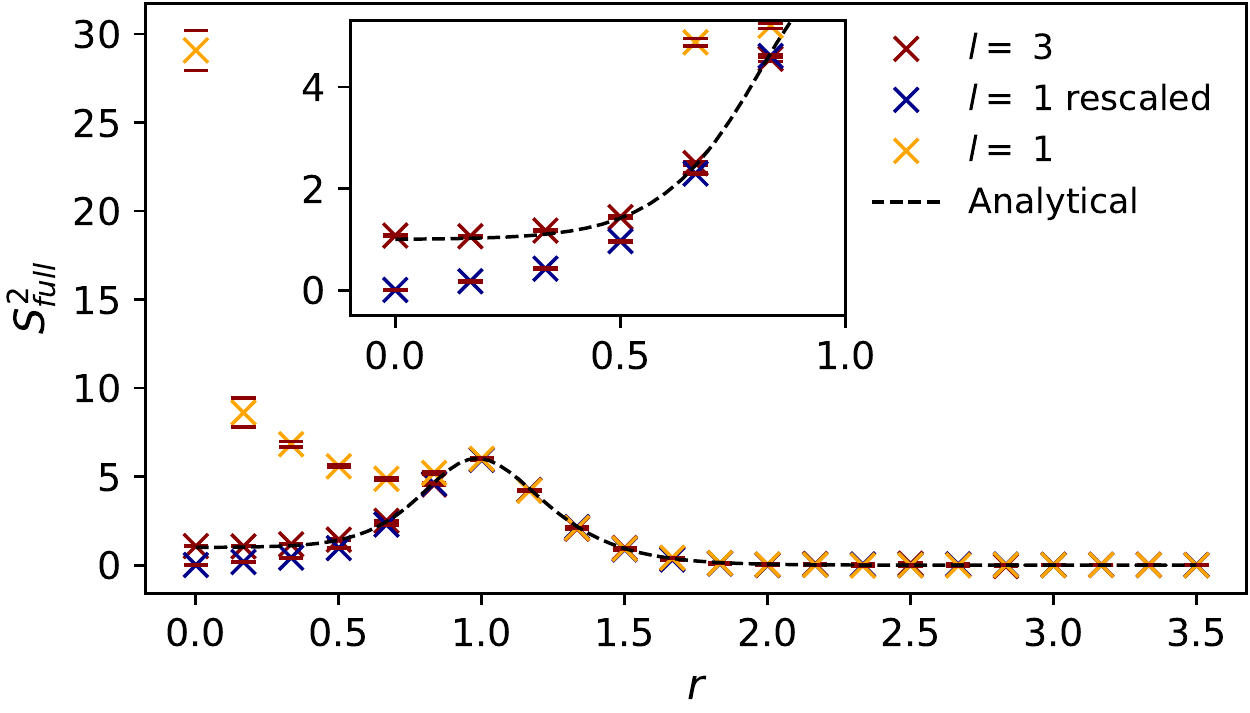}
\caption{\label{Appfig:Appendix_IsingModelPaths}
\miha{Comparison of integration paths $l=3$ and $l=1$ in the Ising model with $64$ spins and $\beta=3$. 
Rescaling $\lambda$
is not sufficient to obtain a reliable result at $l=1$ and path regularisation is needed.}
}
\end{figure}

In 
our
analysis of the Ising model, $\lambda$ was rescaled according to
\begin{eqnarray}
\lambda' = \sin^2(\pi\lambda / 2).
\label{eq:lambda_rescaling}
\end{eqnarray}
This 
helps
to further reduce 
the integration error 
coming from
the unfavorable shape of the integrand (Fig.~\ref{Appfig:Appendix_IsingModel}a). 
We note that the path regularization proposed in the Letter,  
based on a flexible path direction
in the $(g,h)$ Hamiltonian space, is already sufficient to reduce the variance. Rescaling the parameter $\lambda$ according to Eq.~\ref{eq:lambda_rescaling} on top of the path regularization 
can merely be used to further enhance the 
integration efficiency by reducing
the number of integration steps required, as shown in Fig.~\ref{Appfig:Appendix_IsingModel}a. \miha{The need of path regularisation is demonstrated in Fig.~\ref{Appfig:Appendix_IsingModelPaths}, where linear interpolation ($l=1$) is compared with the regularised path. It is also apparent that rescaling 
$\lambda$ according to Eq.~\ref{eq:lambda_rescaling} is less effective than changing the shape, i.e. tuning $l$, in order to regularize the thermodynamic integration. 
}

\section{Formic acid dimer}
\label{formic}

\subsection{Potential energy surface evaluation}
In the formic acid dimer simulations,
we explored the possibility of evaluating quantum entanglement in a truly \emph{ab initio} system. 
It is made of
of two molecules that form a dimer via a double hydrogen bond. 
The R\'enyi entanglement entropy calculation was based upon
the determination of
the potential energy surface (PES) 
for the two protons
involved in the bond.
We 
constrained
them to move only along the hydrogen bond direction, and we fixed the positions of all the other atoms, thus reducing the PES to two dimensions.

\begin{figure}[t]
\includegraphics[width=1\linewidth]{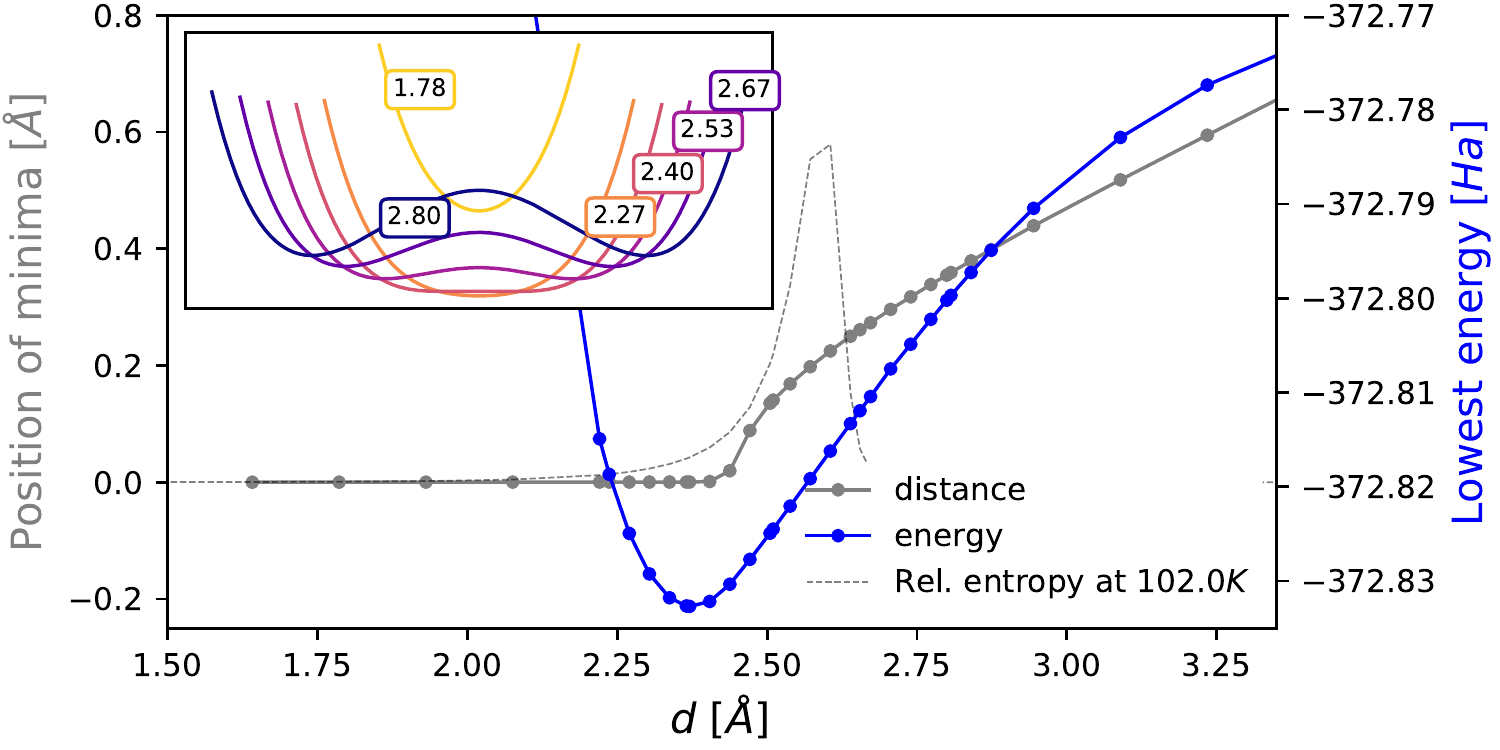}
\caption{\label{Appfig:Appendix_FormicAcid} 
Evolution of the formic acid dimer PES as a function of 
$d$. As $d$ increases, the single minimum splits in two around $d=2.4$ \AA. The distance between the two minima is shown in the main panel (gray points), together with their energy (blue curve). The bifurcation point is very close to the global minimum. The relative entanglement entropy at $102.0$ K is also plotted for comparison (dashed-gray line). The most entangled states appear at distances with higher energy. Inset: PES profile along the diagonal for a few values of $d$.}
\end{figure}

The position of the other atoms was obtained by optimizing the geometry of the system in a symmetric configuration, i.e. at the transition state, where  
the two protons are equally shared. 
This could affect the results at larger distances, but we expect these effects to be small close to the symmetric transition state. After fixing the H-C$<$OO configuration 
for each monomer, only the distance between them 
is
varied 
and, at each distance, the PES 
is
evaluated. The PES constructed in this way has a global minimum close to the one of the transition state (see Fig.~\ref{Appfig:Appendix_FormicAcid}).

The PES was 
computed
with the coupled cluster single double perturbative triple (CCSD(T)) method, using the standard library implemented in the Python-based Simulations of Chemistry Framework (PySCF) package\cite{Sun2018, Sun2020} with the 'STO-3G' basis set.  Because the method is time consuming, the PES is first determined on 
a $30\times30$ points grid,
and then Bspline-interpolated 
on a denser grid of $10,000\times10,000$ points. 

The PES dependence on the distance $d$ between face-to-face oxygen atoms (a proxy for the distance between the two monomers) is shown 
in Fig.~\ref{Appfig:Appendix_FormicAcid}. With increasing 
$d$, the PES minimum 
splits in two, 
and entanglement between the protons drastically increases. The configurations with maximally entangled protons are not the configurations with the lowest energy.

\subsection{Path integral Monte Carlo}
Rényi entanglement entropy was evaluated using two methods: 
Monte Carlo sampling, used to get the main results reported in the Letter, 
and exact diagonalization, used for benchmarking when accessible.

For the simulation of the \textit{ab initio} Hamiltonian only a simple algorithm was used in path integral Monte Carlo, consisting of two types of moves. The first move 
consists
in displacing a randomly chosen pair of beads with a random vector drawn from the uniform distribution of width $\zeta\hbar^2 / (2m)$. The second possible move 
is a mirror transformation, obtained by changing the sign in front of the 
particles position of one copy. The move is always accepted for particles that are not connected, while for the connected particles the only energy difference involved comes from the boundary. This second type of move was necessary for sampling the double well potential, where particles 
get
stuck in one of the wells if the barrier
is
too high. In this case, ergodicity would not be preserved. This additional Monte Carlo move 
was enough to preserve ergodicity 
in
our simulations.

\subsection{Exact diagonalization}
In order to evaluate the entanglement entropy via the exact diagonalization, we used the Hamiltonian truncation method. With this method the infinite Hilbert space is truncated in such a way that the space becomes finite and the low-energy properties stay the same. One can test the correctness of the results by gradually increasing the space and by looking at the eigenvalues. Since the Hilbert space of the system is a product of two one-particle states, $\mathcal{V}_1 \otimes \mathcal{V}_2$, the space can be truncated by introducing only a finite number of possible positions for each particle. In the position space, the kinetic term is represented by finite differences while the potential is diagonal in the basis. We performed calculations on grids spanning from $30\times 30 $ to $100\times100$ points. For the potentials considered, small grids were already extremely accurate.

Once the low-lying eigenvalues are known (at low temperature the high-energy contributions can be neglected), 
the evaluation of the full entropy is straightforward. 
However,
for the entanglement entropy the partial trace has to be performed. Because we know how the space 
is
constructed, the trace can be expressed as
\begin{eqnarray}
\rho_A  =  \sum_i p_{i, i} \sum_{x', x''}\Big(\sum_{y}  c_{i, x', y}c^*_{i, x'', y}\Big)|x'\rangle \langle x''|,
\label{eq:ExactDiagonalization_ReducedDensity}
\end{eqnarray}
where $i$ runs over the eigenstates, $c_{i, x'', y}$ are the coefficients that determine the eigenstates expansion over the position basis set: $|v_i\rangle = \sum_{x}\sum_{y}c_{i, x, y}|x\rangle\otimes |y\rangle$. The partial trace 
requires just the sum over $y$.
After this is computed for each pair $x', x''$, a reduced density matrix is obtained. The remaining object to calculate is 
the trace of $\rho_A$ raised to the power $\alpha$
\begin{eqnarray}
S_A^{\alpha} = \frac{1}{1-\alpha}\log\text{Tr}\rho_A^{\alpha},
\label{Appeq:RenyiEntropyDef}
\end{eqnarray}
which can be done without any diagonalization.



\end{document}